# A New Evolutionary Bayesian Approach Incorporating Additive Path Correction for Nonlinear Inverse Problems

M Venugopal[2], D Roy[1,*] and R M Vasu[2]

[1]Computational Mechanics Lab, Department of Civil Engineering

[2]Department of Instrumentation and Applied Physics

Indian Institute of Science, Bangalore 560012, India

*Corresponding Author; Email: royd@civil.iisc.ernet.in

## Abstract

An evolutionary form of a generalized Bayesian update method, which is strictly derivative-free yet directed through an additive update term based purely on the statistical moments of the design variables, is proposed for nonlinear inverse problems in general and applied in particular to an optical imaging problem, the ultrasound modulated optical tomography (UMOT). The additive update term, which bypasses most pitfalls of a conventional weight-based Bayesian update, results from a change of measures aimed at driving appropriately derived observation-prediction error terms or increments of cost functionals to zero-mean Brownian martingales. This constitutes a novel characterization corresponding to the extremization of the cost functional(s), where the design unknowns are represented as diffusion processes evolving with respect to a continuously parameterized iteration variable. This leads to a recursive prediction-update algorithm to implement the search. The scheme offers freedom from sample degeneracy and the accompanying divergence of the conventional weight-based Bayesian update schemes. We obtain the order of convergence of the conditioned process and also establish that the solutions are stable against tolerable variations in the regularizing noise terms, even as the original inverse problem remains severely ill-posed. Numerical evidence on solutions to the UMOT problem also confirms substantive improvements in the reconstruction efficacy through the proposed method vis-à-vis a Gauss-Newton approach, especially where the regularized quasi-Newton direction has low sensitivity to variations in the design unknowns.

# 1. Introduction

Inverse problems typically aim at recovering the unknown or inadequately known response or parameter fields of a system model, based on noisy and partially procured measurements of the system response. Here the system model (also called the forward model) refers to an appropriate mathematical descriptor, e.g. a family of partial differential equations (PDEs), of the system response under specified inputs. Although inverse problems arise in many areas of science and engineering (e.g. geophysical/medical imaging), most problems of practical importance are ill-posed owing to the compactness of the forward operator as well as the partial noisy measurements that render the solutions for the model parameters non-unique and unstable. The ill-posedness necessitates the use of certain regularization strategies that may include attempts at incorporating some *a priori* information on the model parameters (say, in terms of their smoothness or restriction of the solution space) to yield stable and meaningful solutions. Inverse problems admit formulations that draw upon methods from optimization, wherein the goal is to minimize an objective functional that specifies the misfit between the available measurements and the predictions from the last recovered model. A well known route to arrive at the (ideally global) minimum of the objective functional, leading to the solution of the inverse problem, is the quasi-Newton iteration [1] as exemplified, say, by the Gauss-Newton (GN) algorithm.

We propose an alternative stochastic search approach for a fairly broad class of nonlinear inverse problems, providing for a strictly derivative-free and largely regularization-insensitive numerical scheme. Indeed, the proposed method falls under the broad category of random evolutionary algorithms [2], of which the genetic algorithm (GA) [3] is a well known member. In the current probabilistic setting, the minimization of the measurement-prediction misfit, or the so-called error, that corresponds to the recovery of parameters is achieved by rendering the parameters to be measurable with respect to the filtration generated by the misfit. This is possible once fictitiously introduced and regularizing noise processes, evolving with progressive iterations, are made use of to characterize both the parameters and the misfit as stochastic processes. An alternative interpretation of the above strategy is to derive a change of measures such that, under the new measure, the misfit process becomes a zero-mean martingale, e.g. a Wiener noise process. This is proposed to be accomplished using the following general setting. Let the forward model and the measurement equation be respectively given by:

$$\mathcal{L}(u(\mathbf{x}), p(\mathbf{x})) = \mathrm{f}, \quad \mathbf{x} \in \mathbf{D} \tag{1a}$$

$$\mathcal{M}(u(\mathbf{x})) + \mathbf{\eta}^s = \mathbf{M}, \quad \mathbf{x} \in \partial\mathbf{D} \tag{1b}$$

where $\mathcal{L}$ is the forward operator (not necessarily linear) that takes $p$ and $\mathrm{f}$ as its input and solves for $u$ (the system state or response field) in Eqn. (1a). The measurement operator $\mathcal{M}$ in Eqn. (1b) maps $u(\mathbf{x}) := u(\mathbf{x}, p(\mathbf{x}))$ to the measured quantity $M$ modulo the zero-mean (spatially uncorrelated) noise term $\mathbf{\eta}^s$ at a finite set of points on $\partial\mathbf{D}$ (in some cases, measurements may also be procured from within the domain interior; such a modification may however be readily accommodated within our proposed method). The inverse problem is to recover the parameter field $p(\mathbf{x})$ from the noisy data $\mathbf{M}$ measured on a finite subset of the boundary $\partial\mathbf{D}$. Given a function space setting over which Eqn. (1) is defined, the unknown field $p$ is typically infinite dimensional that needs to be recovered from a finite (and typically small) set of measurements (data), say $\mathbf{M} = \{M^1, M^2, ..., M^{n_M}\}, n_M \in \mathbb{N}$. This often renders the problem severely under-determined. In this context, a stochastic setting (e.g. one based on generalized Bayesian updates, as adopted here) is particularly helpful as the numerous solutions fitting the data may be assigned weights or likelihoods based on prior modelling through continuous probability distributions. In other words, a stochastic setting allows one to specify *a priori* the form of solutions that are deemed more likely and then to define a change of measures that determines the weights. Given the nonlinearity in the forward and/or measurement models, it also follows that any feasible stochastic method, like its deterministic counterparts, may only iteratively guide the predictions based on the forward mathematical model *en route* to the minimization of the measurement-prediction error.

Following a finite-dimensional discretization of Eqn. (1a) through an appropriate scheme (e.g. the finite element method, FEM) and allowing for a convenient abuse of notations, let the discretized parameter vector be written as $\mathbf{p} := \{p^1, p^2, ..., p^{n_p}\}, n_p \in \mathbb{N}$. Being grounded in stochastic modelling, the proposed method treats $\mathbf{p} : \Omega \to \mathbb{R}^{n_p}$ at each iteration as an $n_p$-dimensional, Borel measurable random vector (with its *a priori* distribution drawn, for instance, from the Gaussian family) in an underlying complete probability space $(\Omega, \mathcal{F}, P)$. Here $\mathcal{F}$ may be identified with the Borel $\sigma$-algebra $\mathcal{B}(\mathbb{R}^{n_p})$ over open subsets of $\mathbb{R}^{n_p}$. Within this setting, the posterior distribution for $\mathbf{p}$ could be non-Gaussian and multi-modal, the latter accounting for the multiple solutions that are possible in an underdetermined system as discussed above. Note that the prior measure $Q_0$ induced by $\mathbf{p}$ on $\mathcal{B}(\mathbb{R}^{n_p})$ is a pushforward of the measure $P$, i.e. $Q_0$ may be identified with $P$. A Bayesian solution to the inverse problem that 'matches' the given data may then be viewed as an estimate of $\mathbf{p}$ that is arrived at iteratively by attaching weights to the different realizations based on their corresponding prediction-measurement mismatch. In most existing Bayesian approaches [4, 5, 6, 7], these weights are the likelihood ratios which when multiplied with the prior density form the (empirical) posterior density. Consequently, in subsequent iterations, higher weights will be assigned to the best fit solutions (i.e. those corresponding to smaller errors) and increasingly negligible weights will be assigned to the rest leading to a possible paucity of distinct samples even before the right solution is arrived at.

If $\mathbf{e}$ denotes the misfit between the predictions upon inverting the forward model and the measurements, then $\mathbf{e}$ is an $\mathcal{F}$-measurable random variable and the main idea implemented in this work is to reconstruct the parameter $\mathbf{p}$ as the $\mathcal{F}^e$-measurable estimate $\mathbf{p}^* = E_P\left[\mathbf{p} \mid \mathcal{F}^e\right]$, where $\mathcal{F}^e := \sigma(\mathbf{e})$ is the $\sigma$-algebra generated by the random variable $\mathbf{e}$ and $E_P[.]$ denotes the expectation operator under P. Motivated by the theory of nonlinear stochastic filtering, the present goal is to effect a change of measures $P \rightarrow Q$ such that $E_P\left[\mathbf{p} \mid \mathcal{F}^e\right] = E_Q[\Lambda \mathbf{p}]/E_Q[\Lambda]$, i.e. the conditional expectation is rendered unconditional under Q, where $\Lambda := dP/dQ$ is the Radon-Nikodym derivative. Since Q is not known *a priori* for the nonlinear inverse problem, the transformation $P \rightarrow Q$ needs to be implemented iteratively. Thus define the misfit $\mathbf{e}_k$ at the $k^{\text{th}}$ iteration as:

$$\mathbf{e}_k = \mathbf{M} + \boldsymbol{\eta}_k - \mathcal{M}(\mathbf{u}_k(\mathbf{p}_k)) \qquad (2)$$

$\boldsymbol{\eta}_k$ is a fictitiously applied regularizing noise process (e.g. a random walk or a discrete Brownian motion) in $k$ so that $\boldsymbol{\eta}_0 := \boldsymbol{\eta}^s$ represents the true measurement noise. Indeed, for $k > 0$, $\boldsymbol{\eta}_k$ might also be given the physical interpretation of representing the discretization errors. Let $\mathcal{F}_k^e := \{\sigma(\mathbf{e}_l) : 1 \leq l \leq k\} \cup \sigma(\boldsymbol{\eta}_0)$ denote the increasing filtration generated by the misfit process $\mathbf{e}_l$ up to and including the $k^{\text{th}}$ iteration, with $\sigma(\mathbf{e}_l)$ being the $\sigma$-algebra generated by the random variable $\mathbf{e}_l$ and $\sigma(\boldsymbol{\eta}_0)$ the $\sigma$-algebra corresponding to the true measurement noise. Under this definition, we also have $\mathcal{F}_k^e = \mathcal{F}_k^M$, the latter denoting the filtration generated by the measurement pseudo-process $\mathbf{M}_k = \mathbf{M} + \boldsymbol{\eta}_k$. Note that $k = 0$ corresponds to $\mathbf{p}_0$ being drawn from $Q_0$ and that, under the targeted measure Q, the pseudo-process $\mathbf{M}_k$ should behave as the noise process $\boldsymbol{\eta}_k$ in law. The vector-valued, $\mathcal{F}_k^e$-measurable random variable $\hat{\mathbf{p}}_k = \mathrm{E}_P\left[\mathbf{p} \mid \mathcal{F}_k^e\right]$ obtains the updated parameter at the $k^{th}$ iteration which is input to $\mathcal{L}$ in Eqn. (1a) (or, rather, its discretized form using, say, a finite element method, FEM) to obtain $\mathbf{u}_k$. Here the vector $\mathbf{u}$ denotes the discretized version of the field $u(\mathbf{x})$ and $\mathcal{M}$ corresponds to the measurement operator, defined earlier in Eqn. (1b) and now acting on $\mathbf{u}$ (allowing for a slight notational ambiguity). The iterative change of measure aims at obtaining $Q_k$ such that $\hat{\mathbf{p}}_k = \mathrm{E}_P\left[\mathbf{p} \mid \mathcal{F}_k^e\right] = \mathrm{E}_{Q_k}\left[\Lambda_k \mathbf{p}\right] / \mathrm{E}_{Q_k}\left[\Lambda_k\right]$, where $\Lambda_k$ is the Radon-Nikodym derivative associated with the change of measure. The last equation may be interpreted as the generalized Bayes' formula. A major contribution of this work is in the introduction of an additive correction (or gain) term in place of the (multiplicative) likelihood factor so as to endow $\mathbf{M}_k$ with the same law as $\boldsymbol{\eta}_k$ as $k$ becomes sufficiently large. Indeed, as the intensity of the Brownian motion $\boldsymbol{\eta}_k$, which will be seen to regularize the updating scheme, tends to zero, we may approach the solution available through a strictly deterministic (e.g. the regularized GN) route in principle.

Despite the generality of our setup, numerical exposition of the proposed method is undertaken for a specific inverse problem, which presently is that of ultrasound-modulated optical tomography (UMOT). Some relevant background information on the UMOT, along with a brief account of the methodology to obtain parameter recovery through a GN algorithm, appears in Appendix I. The rest of the paper is organized as follows. The theoretical basis of the proposed method is set forth in Section 2 wherein we also discuss aspects regarding the implementation of the algorithm. We then proceed to establish the stability and convergence of the solutions via the proposed method in Section 3. Specifically, we prove the continuity properties of the solutions (with respect to changes in the data) and their asymptotic convergence to the desired solution. The numerical experiments including details on data generation and reconstruction are in Section 4. The discussion of results and the comparison of the performance of the GN scheme vis-à-vis the proposed method are also given in Section 4. Major conclusions based on this work are summarized in Section 5.

## 2. A Stochastic Search Approach to Inverse Problems

A brief description of the inverse UMOT, a prototypical imaging problem, is given in Appendix I. It may be seen that parameter (i.e., the discretized mean-square displacement vector $\mathbf{p}$) estimation from noisy measurements is set as an inverse problem wherein an objective functional, $\chi(\mathbf{p})$ is sought to be minimized. With most existing formulations, $\chi(\mathbf{p})$ has an error term, supplemented with a regularization term, and the minimization problem is typically stated as:

$$\min_{p \epsilon L^\infty(\mathbf{D})} \chi(\mathbf{p}) = \frac{1}{2} \| \mathbf{F}(\mathbf{p}) - \mathbf{M} \|^2 + \frac{\beta}{2} \| \mathbf{p} \|^2_{L^2(IR)} \qquad (3)$$

where $IR$ denotes the insonified focal region of the ultrasound transducer and $\beta$ is a regularization parameter. Moreover, $\mathbf{F}(\mathbf{p}) := \mathcal{M}(\mathbf{p}, \mathbf{r}, \varpi_a)|_{\mathbf{r} \in \partial \mathbf{D}}$ where the definition of $\mathcal{M}$ as given by Eqn. (A.4) has been slightly modified with $\mathbf{p}$ replacing the field $p$ in the first argument. The minimization is often attempted through a Gauss-Newton (GN) algorithm giving an iterative procedure as $\mathbf{p}_{k+1} = \mathbf{p}_k - \mathbf{H}(\mathbf{p}_k)^{-1} \mathbf{G}(\mathbf{p}_k)$. Here $\mathbf{H}$ and $\mathbf{G}$ respectively denote the Hessian and gradient of $\chi$. They are approximated by $\mathbf{H}(\mathbf{p})(\delta \mathbf{p}) = D\mathbf{F}^*(\mathbf{p}) D\mathbf{F}(\mathbf{p})(\delta \mathbf{p}) + \beta(\delta \mathbf{p})$ and $\mathbf{G}(\mathbf{p}) = D\mathbf{F}^*(\mathbf{p})(\mathbf{F}(\mathbf{p}) - \mathbf{M})$. $D\mathbf{F}$ denotes the Frèchet derivative of $\mathbf{F}$ and $D\mathbf{F}^*$ its adjoint. The details of the implementation of the GN algorithm, including the calculation of the gradient and the Hessian via the adjoint of the forward perturbation equation (A.6), are given elsewhere [8]. The same procedure has been employed here to obtain a recovery of $\mathbf{p}$ so as to serve as a benchmark for the performance of the proposed method described below.

The treatment of an inverse problem within a probabilistic setting invariably entails the unknown solution(s) to be sought within a complete probability space, say $(\Omega, \mathcal{F}, P)$. Since the update procedure must necessarily be iterative, the system parameters to be recovered are appropriately randomized using fictitiously introduced process noises and then evolved, over the iterations, as discrete stochastic processes on $\Omega$. In general, the parameter increments are characterized as incremental random walks (e.g. Weiner or Brownian increments $\Delta \mathbf{B}_k$). As noted earlier, the acquired measurement noise $\boldsymbol{\eta}^s$ may similarly be imparted the structure of a stochastic process by introducing a fictitious noise $\boldsymbol{\eta}_k$ of the random walk type (independent of the process noise added to the parameters) with $\boldsymbol{\eta}_0 = \boldsymbol{\eta}^s$ so that the resulting measurements at the $k^{th}$ iteration are adapted to $\mathcal{F}_k$. Here the filtration $\{\mathcal{F}_k\}$ ($k \in \mathbb{N}$) constitutes an increasing family of sub $\sigma$-algebras of $\mathcal{F}$ containing the history of all the noise processes up to the iteration $k$ and the $\sigma$-algebra generated by the measurement noise $\boldsymbol{\eta}^s$ corresponding to $k = 0$. Continuous and infinitely supported distributions (e.g. Gaussian) of all the noise processes are often useful in a Girsanov change of measures as the distributions involved are absolutely continuous with respect to each other. However, presence of finitely supported constraints in some inverse problems may prevent the use of Brownian motion to represent the artificially introduced noises; possible variations to account for such a scenario are however not addressed in this work. Irrespective of the distribution, the random walk model is a time-homogenous Markov process that enables the associated stochastic states also to be Markovian, a property that characterizes (weak) solutions of stochastic differential equations (SDEs) through transitional probability kernels. The last aspect is crucial in setting up the proposed iterative updates.

*Stochastic characterization of the error function(al) to determine the optimal parameters*:

All (semi-)deterministic formulations for inverse problems, inclusive of the GN approaches, involve extremizing an objective functional, say $\chi(\mathbf{p})$, or minimizing an error functional $|\mathbf{e}(\mathbf{p})|$, where $\mathbf{p}$ corresponds to the model parameter vector to be recovered. For instance, $\chi_k(\mathbf{p})$ could be the sum of the squares of the $n_M$ components of $\mathbf{e}_k$ in the discretized problem corresponding to Eqn. (1), where $\chi_k(.)$ is the $k^{th}$ iterative approximation to $\chi(.)$. As $\mathbf{p}$ approaches its optimal value, say $\mathbf{p}^*$, $\chi(\mathbf{p}^*)$ attains stationarity (in the sense of vanishing first variation about $\mathbf{p}^*$) within a deterministic setting. As a precursor to describing the stochastic search scheme, in what follows, a couple of stochastic versions of the above stationarity condition are proposed.

The stochastic counterpart of the so-called stationarity would be an evolution of $\chi_k(\mathbf{p})$ as a random walk with iteration-invariant mean $\approx \chi_k(\mathbf{p}^*)$ and sufficiently small variance to check the error due to the fictitiously introduced noise processes. It appears possible to characterize the random walk process $\chi_k(\mathbf{p})$ by independent increments over iterations, whose mean should become zero (hence iteration-invariant) as $\mathbf{p}$ approaches $\mathbf{p}^*$ in some sense (say, in the mean square sense). In other words, $\Delta\chi_k(\mathbf{p})$ may be interpreted as behaving as a zero-mean martingale [9] as $k \to \infty$. Such a characterization is consistent with the remarkable property of a martingale in that the expected change in its value is zero over subsequent iterations given the history of its evolution until the present iteration. Indeed, as the variance of the artificial noise processes tends to zero (even though a strictly zero variance is unattainable in the proposed method), the martingale would behave almost as a constant-valued function, the constant value being its mean (the extremal functional value) that corresponds to the optimal parameter. A generalized Bayesian search would then provide an updating scheme for $\mathbf{p}_k$ by driving the incremental objective functional to a martingale. It is now important to verify if the above characterization is rigorously justifiable. For purposes of expositional clarity, consider the simple problem of extremizing the functional:

$$\chi(p) = p^2, p \in \mathbb{R} \tag{4}$$

The optimization problem is to find $p^*$ such that $\chi(p^*) \leq \chi(p) \; \forall p \in \mathbb{R}$. In the stochastic setting, let the variable $p$ evolve as a discrete Brownian motion (with iteration dependent mean) i.e.

$$\Delta p_k = \Delta B_k, \text{ or } p_{k+1} = p_k + \Delta B_k \quad k = 1, 2, \ldots \tag{5}$$

The distribution of the Brownian increment $\Delta B_k$ is zero-mean Gaussian with variance $\Delta \tau_k := \tau_{k+1} - \tau_k$, where $\tau := \tau(\gamma)$ is a strictly increasing function of $\gamma \in \{0\} \cup \mathbb{R}^+$ and $\tau_k = \tau(\gamma = k)$. $\tau$ may be thought of as a time-like parameter so that $\Delta B_k$ may be interpreted as an increment of $B(\tau)$, a continuously parameterized version of the Brownian noise. This in turn also enables continuous parameterizations of $\chi$ and $p$ as $\chi_\tau$ and $p_\tau$ respectively (i.e. we have $\chi_k = \chi_{\tau_k}$, $p_k = p_{\tau_k}$ etc.), thereby enabling one to access the rich theory of stochastic calculus (Ito's calculus, in particular) [9]. In a Monte Carlo setting, let $\{\hat{p}_k(\omega_j) \mid j \in \{1,...,n_E\}\}$ denote the updated ensemble of parameters at the end of the $k^{th}$ iteration, where $n_E$ denotes the ensemble size and $\omega_j \in \Omega$. Using Ito's formula, the incremental objective functional may then be written as

$$\Delta \chi_k = \chi_{k+1} - \chi_k = 2 \int_{\tau_k}^{\tau_{k+1}} p\, dB_\tau + \Delta \tau_k = 2 \int_{\tau_k}^{\tau_{k+1}} (\hat{p}_k + \Delta B_\tau) dB_\tau + \Delta \tau_k = 2\hat{p}_k \Delta B_k + \Delta \tau_k \quad (6a)$$

where $\Delta B_\tau := B_\tau - B_k$ and use has been made of the identities $\int_{\tau_k}^{\tau_{k+1}} \Delta B_\tau dB_\tau = \frac{1}{2}(\Delta B_k^2 - \Delta \tau_k)$ and $\Delta B_k^2 = \Delta \tau_k$ in $L^2(\mathrm{P})$. Now, writing $\chi_k$ as $\chi_k = p_k^2 = (m_k + \Delta B_{k-1})^2$ with $m_k = \mathrm{E}[\hat{p}_k]$, one may rewrite Eqn. (6a) as:

$$\chi_{k+1} = m_k^2 + 2m_k(\Delta B_{k-1} + \Delta B_k) + (\tau_{k+1} - \tau_{k-1}) \quad (6b)$$

Noting that $m_k$ is non-random, we introduce a variation of $m_k$ (and hence in $m_k^2$) and thus write the variation $\delta \chi_{k+1}$ as:

$$\delta \chi_{k+1} = \delta(m_k^2) + 2\delta(m_k)(\Delta B_{k-1} + \Delta B_k) \quad (6c)$$

Note that if $p^* = m_k$ were an extremizer of the deterministic functional $\chi(p) = p^2$ in Eqn. (4), then we would have $\delta(m_k^2) = 0$, thereby reducing the variational process $\delta \chi_{k+1}$ into a $\mathcal{F}_{k+1}$-martingale (as $(\Delta B_{k-1} + \Delta B_k)$ is an $\mathcal{F}_{k+1}$-martingale). Granting that the iterations are so designed as to ensure that $m_k = p^*$ for all $k \geq N_I$ and that $\Delta \tau_k = \Delta \tau$ (a constant) for all $k \geq N_I$, one may refer to Eqn. 6(b) to write:

$$\chi_{k+1} - \chi_k = 2p^*(\Delta B_k - \Delta B_{k-2}); \quad k \geq N_I \quad (7)$$

That is, $\chi_{k+1} - \chi_k$ becomes a zero-mean Wiener martingale with respect to $\mathcal{F}_{k+1}$. However, for computational purposes, a more convenient representation of this martingale would be given by:

$$\chi_{k+1} - \mathrm{E}[\chi_k] = 2p^*(\Delta B_{k-1} + \Delta B_k); \quad \forall k \geq N_I \tag{8}$$

It is of interest to note from Eqn. 6(b) that, as the mean value of $\hat{p}_k$ approaches $p^*$ for $k \geq N_I$ (i.e. as $m_k \to p^*$), the process $\chi_{k+1}$ behaves as an $\mathcal{F}_{k+1}$-submartingale as its mean continues to increase with additional iterations owing to the third term on the RHS (except for the degenerate case when $\tau_k$ becomes non-increasing in $k$, $\forall k \geq N_I$). Clearly, the objective functional in this example is strictly convex with a single (global) minimum. The proposed characterization however admits ready extension to a multi-variable, multimodal objective functional by the fact that any such smooth functional may be approximated locally by a quadratic functional within a neighbourhood of an extremal value. The rest of the derivation remains the same excepting that multi-dimensional version of Ito's formula must be used in writing Eqn. 6(a). Here again, the martingale property of $\chi_{k+1} - \mathrm{E}(\chi_k)$ would ensure that, for sufficiently large $k$, the mean of the parameter-vector process $\mathbf{p}_k$ goes to $\mathbf{p}^*$, an extremizing point for the deterministic functional $\chi(\mathbf{p})$. A generalized Bayesian search method based on this characterization would henceforth be referred to as 'Version-1'.

However, in the context of an inverse problem, Version-1 is not the only (or, even the most desirable) characterization through which the extremization should be achieved. Indeed, given a set of measurements $\mathbf{M} = \{M^1, ..., M^{n_M}\}$ and the associated set of measurement operators $\mathcal{M} = \{\mathcal{M}^1(\mathbf{p}), ..., \mathcal{M}^{n_M}(\mathbf{p})\}$ (omitting the dependence of $\mathcal{M}$ on other arguments), an alternative idea, which is probably more appealing, is to obtain a stochastic search for the optimal parameter vector $\mathbf{p}^*$ so that the observation error set $\mathbf{e}_\tau = \{M_\tau^1 - \mathcal{M}^1(\mathbf{p}_\tau), ..., M_\tau^{n_M} - \mathcal{M}^{n_M}(\mathbf{p}_\tau)\}$ is rendered an $n_M$-dimensional zero-mean martingale (e.g. a Brownian motion) in $\tau$, as defined in describing Version-1. Here $\mathbf{M}_\tau = \mathbf{M} + \mathbf{\eta}_\tau$ denotes the fictitiously created stochastic measurement process in $\tau$. Stochastic search based on such a characterization, which possesses some similarities with nonlinear filtering, will be designated as Version-2. In what follows, a couple of stochastic search methods, based on a change of measures, are described to implement the characterization in Version-2. Note that similar procedures may be derived for Version-1 as well.

*The generalized Bayesian search methods*:

By way of recovering the parameter vector $\mathbf{p}$, we construct the parameter SDEs in $\tau$, valid only over $[\tau_k, \tau_{k+1})$, as follows:

$$d\mathbf{p}_\tau = d\mathbf{B}_\tau \tag{9}$$

In describing our methods, we consider two different forms of the measurement evolution equation. The first of these may be written as:

$$\mathbf{M}_\tau = \mathbf{M}_k + \mathbf{D}\mathbf{K}^{-1}(\mathbf{p}_{k+1})\mathbf{q}\Delta\tau + \Delta\boldsymbol{\eta}_\tau \tag{10a}$$

where $\mathbf{p}_{k+1} \in \mathbb{R}^{n_p}$ is the predicted parameter vector at $\tau_{k+1}$ through Eqn. (9) and $\mathbf{B}_\tau \in \mathbb{R}^{n_p}$ is a Brownian process with mean zero and covariance matrix $\Sigma_B \Sigma_B^T \in \mathbb{R}^{n_p \times n_p}$. $\mathbf{u} = \mathbf{K}^{-1}(\mathbf{p})\mathbf{q}$ obtains the discretized solution of the forward model given by Eqn. 1(a), with $\mathbf{q} \in \mathbb{R}^{n_q}$ denoting the discretized source vector (e.g. corresponding to the source term f in Eqn. 1(a) or the term on the right hand side of Eqn. A6 for the UMOT problem; see Appendix I). $\mathbf{K}(\mathbf{p}) \in \mathbb{R}^{n_q \times n_q}$ is the coefficient matrix obtained by the FEM-based discretization of Eqn. 1(a) and it could be nonlinear in the parameter vector $\mathbf{p}$. Also, $\mathbf{M}_\tau \in \mathbb{R}^{n_M}$ defines the $\tau$-evolving measurement process, given in its algebraic form and including the fictitiously applied regularizing Brownian noise $\boldsymbol{\eta}_\tau \in \mathbb{R}^{n_M}$ with $\boldsymbol{\eta}_0 := \boldsymbol{\eta}^s$ (the latter being the statically acquired true noise), and $\mathbf{D} \in \mathbb{R}^{n_M \times n_q}$ is a binary coefficient matrix used to retrieve the measurements at the appropriate nodes (located on the domain boundary for the UMOT problem). Here $\boldsymbol{\eta}_\tau - \boldsymbol{\eta}^s$ is a vector Brownian motion with zero mean and covariance matrix $\Sigma_\eta \Sigma_\eta^T \in \mathbb{R}^{n_M \times n_M}$. The advantage of writing $\mathbf{M}_\tau$ as in Eqn. (10a) is that it may be readily recast in an SDE form, which is given by:

$$d\mathbf{M}_\tau = \mathbf{D}\mathbf{K}^{-1}(\mathbf{p}_{k+1})\mathbf{q}d\tau + d\boldsymbol{\eta}_\tau \tag{10b}$$

where $\tau \in [\tau_k, \tau_{k+1})$ and $\mathbf{M}_0 := \mathbf{M}(\tau = 0)$ obtains the true data. The second form of the measurement evolution equation is purely algebraic (without a need to be recast as an SDE) and is given by:

$$\mathbf{M}_{k+1} = \mathbf{DK}^{-1}(\mathbf{p}_{k+1})\mathbf{q} + \mathbf{\eta}_{k+1} \tag{10c}$$

As indicated in Eqn. (9) $\mathbf{B}_\tau$ imparts the character of a non-zero mean (local) Brownian martingale to the artificial evolution of $\mathbf{p}_\tau$; however Eqn. (9) does not, by itself, yield any drift information over the iterations. Given an updated ensemble $\{\hat{\mathbf{p}}_k(j) := \hat{\mathbf{p}}_k(\omega_j)\}_{j=1}^{n_E}$ of parameter realizations at the end of the $k^{th}$ iteration ($k = 1, 2, ...$), our iterative method aims at establishing a prediction-update strategy that will first generate the $(k+1)^{th}$ prediction ensemble $\{\mathbf{p}_{k+1}(j) := \mathbf{p}_{k+1}(\omega_j)\}_{j=1}^{n_E}$ using $\{\hat{\mathbf{p}}_k(j)\}_{j=1}^{n_E}$ as the initial conditions. Following this, based on a change of measure, the updated ensemble $\{\hat{\mathbf{p}}_{k+1}(j)\}_{j=1}^{n_E}$ will be determined through an appropriate correction term additively applied to $\{\mathbf{p}_{k+1}(j)\}$ so as to bridge the measurement-prediction misfit. Considering, for instance, the measurement equation (10a) or its SDE-form (10b), the misfit may be characterized through the incremental error represented as $\Delta \mathbf{e}_k = \mathbf{M}_{k+1} - \mathbf{M}_k - \mathbf{DK}^{-1}(\mathbf{p}_{k+1})\mathbf{q}\Delta\tau_k$. Note that the incremental error process $\Delta \mathbf{e}_\tau$ should ideally behave as $\Delta\mathbf{\eta}_\tau$, a Brownian increment with zero mean. On the other hand, if the measurement equation (10c) were considered, the misfit would be given by $\mathbf{e}_{k+1} = \mathbf{M}_{k+1} - \mathbf{DK}^{-1}(\mathbf{p}_{k+1})\mathbf{q}$. In a typical Bayesian approach applied to the current setup (i.e. with the measurement equation (10c)), the update process would involve finding the posterior densities via the product measure defined by

$$\rho_\eta^M(\hat{\mathbf{p}}_{k+1}(j)) = \frac{1}{W}\rho_\eta(\mathbf{e}_{k+1}(j))\rho_B(\mathbf{p}_{k+1}(j)) \tag{11a}$$

where $\rho_\eta$ and $\rho_B$ denote the densities (with respect to the Lebesgue measure) of the Brownian processes $\mathbf{\eta}_\tau$ and $\mathbf{B}_\tau$ respectively. Assuming that the normalizing constant $W = \int_{\mathbb{R}^{n_p}} \rho_\eta(\mathbf{e}_{k+1})\rho_B(\mathbf{p}_{k+1})d\mathbf{p}_{k+1} > 0$, $\rho_\eta^M(\hat{\mathbf{p}}_{k+1})$ is the posterior density, whose mean should yield the optimal solution $\mathbf{p}^*$ as $k \to \infty$, and $\rho_\eta(\mathbf{e}_{k+1})$ is the likelihood. Letting $\mu_\eta^M$ and $\mu_B$ to be measures on $\mathbb{R}^{n_p}$ with densities $\rho_\eta^M$ and $\rho_B$ respectively and assuming that the posterior is absolutely continuous with respect to the prior, the likelihood is proportional to the Radon-Nikodym derivative, which may be written as:

$$\frac{d\mu_\eta^M}{d\mu_B}(\mathbf{p}_{k+1}(j)) = \frac{1}{W}\rho_\eta(\mathbf{e}_{k+1}(j)) \qquad (11b)$$

Specifically, then, the $j^{th}$ updated realization (particle) of the parameter will be given by the pair $(\mathbf{p}_{k+1}(j), w(j))$, where the $j^{th}$ weight is given by $w(j) = \frac{d\mu_\eta^M}{d\mu_B}(\mathbf{p}_{k+1}(j))$. The updates $\{\hat{\mathbf{p}}_{k+1}(j)\}$ may now be generated by resampling over $(\mathbf{p}_{k+1}(j), w(j))$ so that each updated realization has the same weight $(1/n_E)$. This yields the updated approximation to the optimal parameter as $\bar{\hat{\mathbf{p}}}_{k+1,n_E} = (1/n_E)\sum_{j=1}^{n_E}\hat{\mathbf{p}}_{k+1}(j)$ at the end of the $(k+1)^{th}$ iteration. However, the weight based approach as above suffers from degeneracy, i.e. many realizations might receive less weight if the significant mass of the likelihood occupies a small region in $\mathbb{R}^{n_p}$ where the prior density $\rho_B$ is significant. The problem is still aggravated if the likelihood falls in a region of low prior density leading to wastage of realizations distant from the likelihood. It has been shown in [10] that the typical ensemble size preventing such degeneracy is computationally prohibitive with increasing system dimension. In the description to follow, a novel way of effectively tackling this problem is suggested wherein the measurement-prediction error is driven to a zero-mean martingale by additively updating the predictions via a Girsanov transformation of measures [9].

*Scheme 1: Kushner-Stratonovich (KS) additive updates:*

The measurement equation applicable to this scheme would be given by Eqn. (10a) or its SDE-form (10b). Let $Q$ be the probability measure on $(\Omega, \mathcal{F})$ such that the misfit $\Delta \mathbf{M}_\tau$ is a $\tau$-Brownian increment on $(\Omega, \mathcal{F}, Q)$. In this scheme $\mathcal{F}_\tau^e$ denotes the union of the filtration generated by $\Delta \mathbf{e}_s$ for $0 < s \leq \tau$ and the sigma-algebra $\sigma(\mathbf{\eta}^s)$. Assuming that

$$\{(\mathbf{DK}^{-1}(\mathbf{p}_{k+1})\mathbf{q})^i\}^2(\tau - \tau_k) < \infty \ \forall i = 1, 2, ..., n_M, \tau \in [\tau_k, \tau_{k+1}), \text{ P a.s.,}$$

where $(.)^i$ represents the $i^{th}$ component of the vector $\left(\mathbf{DK}^{-1}(\mathbf{p}_{k+1})\mathbf{q}\right)$, and that the measures P and Q are equivalent, i.e. P<<Q and Q<<P, the Radon-Nikodym derivative for $\tau \in [\tau_k, \tau_{k+1})$ is given by:

$$\Lambda_\tau = \left(\frac{d\mathrm{P}}{d\mathrm{Q}}\right)_\tau = \exp\left(\left(\mathbf{DK}^{-1}(\mathbf{p}_{k+1})\mathbf{q}\right)^T (\mathbf{M}_\tau - \mathbf{M}_k) - \frac{1}{2}\left(\mathbf{DK}^{-1}(\mathbf{p}_{k+1})\mathbf{q}\right)^T \left(\mathbf{DK}^{-1}(\mathbf{p}_{k+1})\mathbf{q}\right)(\tau - \tau_k)\right)$$
(12)

$\Lambda_\tau$ is an $\mathcal{F}_\tau$- martingale of the exponential type. An estimate $\bar{\hat{\mathbf{p}}}_{k+1}$ of the optimal solution at the $(k+1)^{th}$ iteration (i.e. corresponding to $\tau = \tau_{k+1}$) is obtained by multiplying the predictions with the Radon-Nikodym derivative and this may be represented via the generalized Bayesian rule as:

$$\bar{\hat{\mathbf{p}}}_{k+1} := \mathrm{E}_\mathrm{P}\left[\mathbf{p}_{k+1} \mid \mathcal{F}_{k+1}^e\right] = \frac{\mathrm{E}_\mathrm{Q}\left[\mathbf{p}_{k+1}\Lambda_{k+1}\right]}{\mathrm{E}_\mathrm{Q}\left[\Lambda_{k+1}\right]}$$
(13)

where the expectation under Q can be evaluated unconditionally as $\Delta \mathbf{M}_k$ is rendered a (discrete) Brownian increment. Now, in view of a possible degeneracy (as discussed earlier) in a weight based approach, we prefer to get an additive update to obtain $\hat{\mathbf{p}}_{k+1}$ from $\mathbf{p}_{k+1}$ based on the above change of measure. Specifically, we arrive at a nonlinear gain-like correction term, which is functionally analogous to the weight and when added to the predicted realization gives the updated realization. This update may be derived by expanding $\hat{\mathbf{p}}_\tau = \mathbf{p}_\tau \Lambda_\tau$ using Ito's formula and, for expositional clarity, its derivation would be explicitly demonstrated for $n_p = 1, n_M = 1$, i.e. $p_\tau := p_\tau^1$ and $dM_\tau = \mathcal{M}(p_\tau)d\tau + d\eta_\tau$ and $\mathcal{M}(p_\tau) = \left(\mathbf{DK}^{-1}(p_\tau).\mathbf{q}\right)$, where $\mathbf{D}$ is an appropriate vector of binary {0,1} entries (note that, in most cases, we would be replacing $p_\tau$ in the argument of $\mathcal{M}$ by $p_{k+1}$ for $\tau \in [\tau_k, \tau_{k+1})$). The corresponding Radon-Nikodym derivative may be written as

$$\Lambda_\tau = \left(\frac{d\mathrm{P}}{d\mathrm{Q}}\right)_\tau = \exp\left(\int_{\tau_k}^\tau \mathcal{M}(p_s)dM_s - \frac{1}{2}\int_{\tau_k}^\tau \mathcal{M}^2(p_s)ds\right)$$
(14)

Now, expanding $\hat{p}_\tau = p_\tau \Lambda_\tau$ for $\tau \in [\tau_k, \tau_{k+1})$ using Ito's formula, we obtain

$$d(p_\tau \Lambda_\tau) = p_\tau d\Lambda_\tau + dp_\tau \Lambda_\tau + dp_\tau d\Lambda_\tau \tag{13a}$$

By replacing $dp_\tau = dB_\tau$ and $d\Lambda_\tau = \Lambda_\tau \mathcal{M}(p_\tau)dM_\tau$ in Eqn. (13a), we get

$$d(p_\tau \Lambda_\tau) = p_\tau \Lambda_\tau \mathcal{M}(p_\tau)dM_\tau + \Lambda_\tau dB_\tau + \Lambda_\tau \mathcal{M}(p_\tau)dM_\tau dB_\tau \tag{13b}$$

The term containing $dM_\tau dB_\tau$ is zero since $M_\tau$ is independent of $B_\tau$ resulting in the following integral representation:

$$p_{k+1}\Lambda_{k+1} = p_k \Lambda_k + \int_{\tau_k}^{\tau_{k+1}} p_s \Lambda_s \mathcal{M}(p_s)dM_s + \int_{\tau_k}^{\tau_{k+1}} \Lambda_s dB_s \tag{14}$$

where $(.)_k = (.)_{\tau_k}$. The unnormalized conditioned estimate $\pi_u(p_\tau) := \pi_{u,\tau}(p) = E_Q[p_\tau \Lambda_\tau]$ may be thus arrived at via the following evolution equation (an equivalent of the Zakai equation in nonlinear filtering theory):

$$\pi_{u,k+1}(p) = \pi_{u,k}(p) + \int_{\tau_k}^{\tau_{k+1}} \pi_{u,s}[p\mathcal{M}(p)]dM_s \tag{15a}$$

To obtain the normalized estimate $\bar{p} = \pi_\tau(p) = \pi(p_\tau) = \dfrac{\pi_{u,\tau}(p)}{\pi_{u,\tau}(1)}$, we write Eqn. (15a) in the incremental form as:

$$d\pi_{u,\tau}(p) = \pi_{u,\tau}(p\mathcal{M}(p))dM_\tau \tag{15b}$$

which yields $d\pi_{u,\tau}(1) = \pi_{u,\tau}(\mathcal{M}(p))dM_\tau = \pi_{u,\tau}(1)\pi_\tau(\mathcal{M}(p))dM_\tau$. Also, we have

$$d\left(\frac{1}{\pi_{u,\tau}(1)}\right) = -\frac{1}{\pi_{u,\tau}^2(1)}d\pi_{u,\tau}(1) + \frac{1}{\pi_{u,\tau}^3(1)}d\pi_{u,\tau}(1)d\pi_{u,\tau}(1) = -\frac{\pi_\tau(\mathcal{M}(p))}{\pi_{u,\tau}(1)}dM_\tau + \frac{\pi_\tau^2(\mathcal{M}(p))}{\pi_{u,\tau}(1)}d\tau$$

$$\tag{16}$$

Using integration by parts and replacing the expressions for $d\pi_{u,\tau}(p)$ and $d\left(\dfrac{1}{\pi_{u,\tau}(1)}\right)$, we get

$$d\pi_\tau(p) = \frac{d\pi_{u,\tau}(p)}{\pi_{u,\tau}(1)} + \pi_{u,\tau}(p)d\left(\frac{1}{\pi_{u,\tau}(1)}\right) + d\pi_{u,\tau}(p)d\left(\frac{1}{\pi_{u,\tau}(1)}\right) \quad (17a)$$

$$d\pi_\tau(p) = \frac{\pi_{u,\tau}(p\mathcal{M}(p))}{\pi_{u,\tau}(1)}dM_\tau - \frac{\pi_{u,\tau}(p)}{\pi_{u,\tau}(1)}\pi_\tau(\mathcal{M}(p))dM_\tau + \frac{\pi_{u,\tau}(p)}{\pi_{u,\tau}(1)}\pi_\tau^2(\mathcal{M}(p))d\tau$$
$$- \frac{\pi_{u,\tau}(p\mathcal{M}(p))}{\pi_{u,\tau}(1)}dM_\tau \pi_\tau(\mathcal{M}(p))dM_\tau \quad (17b)$$

After replacing $dM_\tau dM_\tau \stackrel{L^2(P)}{=} d\tau$, one has:

$$d\pi_\tau(p) = \{\pi_\tau(p\mathcal{M}(p)) - \pi_\tau(p)\pi_\tau(\mathcal{M}(p))\}dM_\tau + \{\pi_\tau(p)\pi_\tau^2(\mathcal{M}(p)) - \pi_\tau(p\mathcal{M}(p))\pi_\tau(\mathcal{M}(p))\}d\tau \quad (17c)$$

Eqn. (17c) may be rearranged to get:

$$d\pi_\tau(p) = d\pi(p_\tau) = \{\pi_\tau(p\mathcal{M}(p)) - \pi_\tau(p)\pi_\tau(\mathcal{M}(p))\}\{dM_\tau - \pi_\tau(\mathcal{M}(p))d\tau\} \quad (18)$$

Note that the above equation is similar to the Kushner-Stratonovich (KS) equation in nonlinear filtering [11]. The term $\{\pi_\tau(p\mathcal{M}(p)) - \pi_\tau(p)\pi_\tau(\mathcal{M}(p))\}$ may be interpreted as the gain-like term that updates the prediction process $p_\tau$ so as to drive $M_\tau - \int_{\tau_k}^{\tau}\pi_s(\mathcal{M}(p))ds$ to a Brownian motion. The evolution of the conditioned process $\pi_\tau(p)$ may be obtained, in principle, by solving the KS-type equation (Eqn. (18)). Unfortunately an exact solution is infeasible except under conditions of linearity of $\mathcal{M}$ and the Gaussianity of all the associated noises [12]. Due to the inherent circularity associated with the moment closure problem, solving this equation is highly non-trivial whenever the function $\mathcal{M}$ is nonlinear, even if the regularizing noise terms are assumed to be Gaussian. We therefore adopt a Monte-Carlo approach to solve the KS-type equation [13]. First, in the more realistic context of higher dimensional inverse problems, the multidimensional (and recursive) variant of Eqn. (18) is written component-wise and in the integral form as

$$\pi_{k+1}(p^l) = \pi_k(p^l) + \sum_{d=1}^{n_M}\int_{\tau_k}^{\tau_{k+1}}\{\pi_s(p^l\mathcal{M}^d(\mathbf{p})) - \pi_s(p^l)\pi_s(\mathcal{M}^d(\mathbf{p}))\}\{dM_s^d - \pi_s(\mathcal{M}^d(\mathbf{p}))ds\}$$
$$\forall l = 1, 2, ..., n_p$$

(19)

Recall that $\mathbf{p} = \left( p^1, ..., p^{n_p} \right)$, $\mathcal{M}(\mathbf{p}) = \left( \mathbf{DK}^{-1}(\mathbf{p})\mathbf{q} \right)$ and $\mathcal{M}(.) = \left( \mathcal{M}^1(.), ..., \mathcal{M}^{n_M}(.) \right)$. Based on Eqn. (19), the iterative equation yielding the set of updated realizations $\{\hat{\mathbf{p}}_{k+1}(j)\}_{j=1}^{n_E}$ corresponding to the $(k+1)^{\text{th}}$ iterate is given by:

$$\hat{p}_{k+1}^l(j) = p_{k+1}^l(j) + \sum_{d=1}^{n_M} \left\{ \hat{\pi}_{k+1}^d \left( p^l \mathcal{M}^d(\mathbf{p}) \right) - \hat{\pi}_{k+1}^d \left( p^l \right) \hat{\pi}_{k+1}^d \left( \mathcal{M}^d(\mathbf{p}) \right) \right\} \left\{ \Delta M_{k+1}^d - \left( \mathcal{M}^d \left( \mathbf{p}_{k+1}(j) \right) \right) \Delta \tau_k \right\}$$

$$\forall l = 1, 2, ..., n_p, \; j = 1, 2, ..., n_E$$

(20)

where $\Delta M_{k+1}^d = M_{k+1}^d - M_k^d$, $\Delta \tau_k = \tau_{k+1} - \tau_k$ and $\{\hat{\mathbf{p}}_{k+1}(j)\}$ is the updated parameter realization set following the $(k+1)^{\text{th}}$ iteration. Obtaining the estimated parameter $\hat{\pi}_{k+1}^d(\mathbf{p}) = \bar{\hat{\mathbf{p}}}_{k+1,n_E} = \frac{1}{n_E} \sum_{j=1}^{n_E} \hat{\mathbf{p}}_{k+1}(j)$ involves the ensemble mean operation and the superscript $d$ denotes the discretization in $\tau$. Note that, in Eqn. (20), $\{\mathbf{p}_{k+1}(j)\}$ is the set of predicted parameters based on the purely diffusive approximation in Eqn. (9), i.e. $\mathbf{p}_{k+1}(j) = \hat{\mathbf{p}}_k(j) + \Delta \mathbf{B}_k(j)$, where $\Delta \mathbf{B}_k(j) = \mathbf{B}_{k+1}(j) - \mathbf{B}_k(j)$. Also note that the ensemble mean operation, although defined above for the updated particles, is similarly used for obtaining the predicted mean $\bar{\mathbf{p}}_{k+1,n_E}$ whenever necessary. The purpose of using the ensemble mean operator is clearly specified wherever there is scope for ambiguity.

*Scheme 2: Least squares (LS) additive updates:*

In this scheme, we make use of a least squares gain (LSG) in place of the KS gain (KSG) to update the parameter realizations where the algebraic form of the measurement evolution equation (10c) is employed. The derivation of the LSG parallels that of the ensemble Kalman gain in [14], wherein a nonlinear filtering problem is solved under the assumption that the associated noises are Gaussian. Here, the predicted realizations $\{\mathbf{p}_{k+1}(j)\}_{j=1}^{n_E}$ are updated using a gain matrix $\mathbf{G}_{k+1}$ as follows:

$$\hat{\mathbf{p}}_{k+1}(j) = \mathbf{p}_{k+1}(j) + \mathbf{G}_{k+1} \left( \mathbf{M}_{k+1} - \mathcal{M}(\mathbf{p}_{k+1}(j)) \right) \tag{21}$$

where $\mathbf{G}_{k+1} = \mathbf{P}_{k+1} \mathbf{M}_{k+1}^T \left( \mathbf{M}_{k+1} \mathbf{M}_{k+1}^T + \Sigma_\eta \Sigma_\eta^T \right)^{-1}$. $\mathbf{P}_{k+1}$ and $\mathbf{M}_{k+1}$ are perturbation matrices given by

$$\mathbf{P}_{k+1} = \frac{1}{\sqrt{n_E - 1}}[\mathbf{p}_{k+1}(1) - \frac{1}{n_E}\sum_{j=1}^{n_E}\mathbf{p}_{k+1}(j),...,\mathbf{p}_{k+1}(n_E) - \frac{1}{n_E}\sum_{j=1}^{n_E}\mathbf{p}_{k+1}(j)]$$

and $\mathbf{M}_{k+1} = \frac{1}{\sqrt{n_E - 1}}[\mathcal{M}(\mathbf{p}_{k+1}(1)) - \frac{1}{n_E}\sum_{j=1}^{n_E}\mathcal{M}(\mathbf{p}_{k+1}(j)),...,\mathcal{M}(\mathbf{p}_{k+1}(n_E)) - \frac{1}{n_E}\sum_{j=1}^{n_E}\mathcal{M}(\mathbf{p}_{k+1}(j))]$

For an effective global search, it is also observed that employing a scalar diffusion factor with the gain-based correction term (in the KSG or LSG scheme) may improve the results by forcing out the realizations stuck around local extrema. Indeed, this multiplicative factor ($(1+\alpha_{k+1})$) may be envisaged as an annealing-like parameter that is added to boost the mixing property of the associated transition kernel. If $\alpha$ denotes the added diffusion parameter, the modified equations corresponding to the KSG and LSG schemes may be written respectively as:

$$\hat{p}_{k+1}^l(j) = p_{k+1}^l(j) +$$
$$(1+\alpha_{k+1})\sum_{d=1}^{n_M}\left\{\hat{\pi}_{k+1}^d\left(p^l\mathcal{M}^d(\mathbf{p})\right) - \hat{\pi}_{k+1}^d\left(p^l\right)\hat{\pi}_{k+1}^d\left(\mathcal{M}^d(\mathbf{p})\right)\right\}\left\{\Delta M_{k+1}^d - \left(\mathcal{M}^d(\mathbf{p}_{k+1}(j))\right)\Delta\tau_k\right\}$$
$$\forall l = 1, 2,...,n_p, j = 1, 2,...,n_E$$

(22)

$$\hat{\mathbf{p}}_{k+1}(j) = \mathbf{p}_{k+1}(j) + (1+\alpha_{k+1})\mathbf{G}_{k+1}\left(\mathbf{M}_{k+1} - \mathcal{M}(\mathbf{p}_{k+1}(j))\right) \tag{23}$$

The added diffusion parameter is functionally analogous to temperature in a simulated annealing scheme [15]. However, we employ a non-conventional schedule for α with an exponential decay [13], $\alpha_{k+1} = \frac{\alpha_k}{\exp(k)}$ to iteratively reduce $\alpha$ to zero. Unlike a Markov chain, the proposed schemes follow ensemble-based simulations, which in itself provides for a reasonably effective exploration of the state space thereby allowing $\alpha_k$ to be reduced quite sharply over the successive iterations.

Finally, it is of interest to observe that the change of measure leading to an additive update procedure is general enough to be adapted to any martingale characterization (e.g. a Poisson martingale or any other non-Brownian martingale) of the observation-prediction error. For instance, the KS- or LS-gain matrix based on the change of measure may be readily computed to drive the augmented error set $\mathbf{e}_{k+1} = \{\chi_{k+1} - E(\chi_k), M_{k+1}^{(1)} - \mathcal{M}^1(\mathbf{p}_{k+1}), ..., M_{k+1}^{n_M} - \mathcal{M}^{n_M}(\mathbf{p}_{k+1})\}$ to a zero-mean $(n_M + 1)$-dimensional zero-mean martingale, where $\{\chi_k = (\mathbf{M}_k - \mathcal{M}_k)^T(\mathbf{M}_k - \mathcal{M}_k)\}$ is the set of error terms. As yet another alternative, one may define the squared error vector $\chi_k = \{\chi_k^d\} = \{(M_k^d - \mathcal{M}_k^d)^2\}$, $d = 1,...,n_M$, and hence the error set

$\mathbf{e}_{k+1} = \{\chi_{k+1}^1 - E(\chi_k^1), ..., \chi_{k+1}^{n_M} - E(\chi_k^{n_M}), M_{k+1}^1 - \mathcal{M}^1(\mathbf{p}_{k+1}), ..., M_{k+1}^{n_M} - \mathcal{M}^{n_M}(\mathbf{p}_{k+1})\}$

which, in turn, could be driven to a zero-mean $2n_M$-dimensional zero-mean martingale. Indeed, part of the strength of the new method should arise from the possibility of its non-unique design against a given problem.

## 3. Sensitivity and Convergence

Having discussed the proposed method in detail, we now explore specific characteristics of the conditioned process, which is evolved over a continuous time-like parameter $\tau$ to recover the model parameters. Note that, subject to the Lipschitz continuity of all the drift and diffusion coefficients, the existence and uniqueness of the conditioned process is guaranteed via the theorems on the 'martingale problem', proved in the celebrated work of Stroock and Varadhan. The work in this section is focussed only on KS-based Scheme 1, even though a similar study could be undertaken for Scheme 2 also. The parameterization in $\tau$ enables the continuous representation of the regularizing Brownian motion processes and this was observed to be helpful in the context of minimizing a given functional. Specifically, in subsection 3.1 we establish a form of well-posedness of the algorithm by proving that the measurements are continuously dependent on the recovered model parameters.

We employ two kinds of approximations in the gain-based update strategy for estimating the conditioned parameter process. The first approximation involves discretization in $\tau$ leading, for instance, to Eqn. (19) that describes the evolution of the $\tau$-discretized conditioned process. Secondly, the iteration dependent law of the parameter vector that should ideally converge to a posterior measure on the recovered parameter, say $\mathbf{p}^*$, is approximated by an empirical measure that is described by a set of $n_E$ particles. The two approximations (viz. the $\tau$-discretization and the particle approximation) are combined in Eqn. (20) to arrive at a numerically feasible recursive algorithm that uses the KSG for updating the parameters.

However it is convenient to separate the errors due to the two approximations to study the convergence via the KSG scheme. A convergence study is presented in subsection 3.2.

**3.1 Well-posedness**

Solutions to inverse problems using most conventional methods are prone to fluctuations owing to small changes in the measurement. It is therefore important to establish the well-posedness of the numerical solution, i.e. to prove that variations in the reconstructed solutions (via the KSG scheme) do not sensitively depend on small variations in the measurement noise. In [16], a form of well-posedness is proved for the conventional (weight-based) Bayesian approach for inverse problems. For expositional convenience, we presently prove the well-posedness in the case of $n_p = 1$ and $n_M = 1$. Suppose that $m$ and $m'$ are two realizations of the 'true' measurement $M_0$ at $\tau_0$ such that $|m-m'|<\varepsilon$, $\varepsilon > 0$. We take $\tau_k \gg \tau_1$ so that for $\tau \geq \tau_k$ the evolving parameters may be deemed to correspond to the finally reconstructed ones. The aim is then to establish that the solution $\pi_k(p)$ is 'stable' in the sense that it is insensitive (or only tolerably sensitive) to perturbations in the measurement.

As noted during the derivation of the KSG, reconstructions are arrived at by conditioning the parameter processes on the filtration generated by the pseudo-measurement processes (or, equivalently, the associated error processes). Hence, corresponding to the two measurements $m$ and $m'$, we obtain two estimates which may be denoted as $\pi_k(p) := E_P\left[ p_k | \mathcal{F}_k^M \right]$ and $\pi'_k(p) := E_{P'}\left[ p_k | \mathcal{F}_k^{M'} \right]$ respectively. Here $M_k$ and $M'_k$ are the two pseudo-measurement processes (starting with $m$ and $m'$ as the initial conditions) obtaining the respective sequences $\{m, M_1, ..., M_k, ...\}$ and $\{m', M'_1, ..., M'_k, ...\}$ in discrete $\tau$. Also, P and P' are arrived at (in terms of their approximate empirical distributions) by changes of measures, the Radon-Nikodym derivatives of which are respectively given by:

$$\left(\frac{dP}{dQ}\right)_k = \Lambda_k = \exp\left(\int_0^{\tau_k} \mathcal{M}(p_s) dM_s - \frac{1}{2}\int_0^{\tau_k} \mathcal{M}^2(p_s) ds\right) \qquad (24a)$$

$$\left(\frac{dP'}{dQ}\right)_k = \Lambda'_k = \exp\left(\int_0^{\tau_k} \mathcal{M}(p'_s) dM'_s - \frac{1}{2}\int_0^{\tau_k} \mathcal{M}^2(p'_s) ds\right) \qquad (24b)$$

Recall that the measurement $M_k$ (or rather $M_\tau$) is a Brownian motion with respect to $Q$. The integrals in Eqns. (24a) and (24b) may be approximated using the Euler-Maruyama scheme to give the following expressions for $\Lambda_k$ and $\Lambda'_k$:

$$\Lambda_k \approx \exp\left(\sum_{i=0}^{k-1} \mathcal{M}(p_i)\Delta M_i - \frac{1}{2}\sum_{i=0}^{k-1} \mathcal{M}^2(p_i)\Delta\tau\right)$$

$$\Lambda'_k \approx \exp\left(\sum_{i=0}^{k-1} \mathcal{M}(p'_i)\Delta M'_i - \frac{1}{2}\sum_{i=0}^{k-1} \mathcal{M}^2(p'_i)\Delta\tau\right)$$

where, following the notations used earlier, $p_i$ corresponds to the prediction at $\tau_i$ and $\Delta M_i = M_{i+1} - M_i$. The two predicted processes $p_i$ and $p'_i$ result from the initial conditions $m$ and $m'$ respectively. In order to quantify the distance between the estimates $\pi_k(p)$ and $\pi'_k(p)$, we make use of the Hellinger metric on the measures $P$ and $P'$. Given that $P$ and $P'$ are absolutely continuous with respect to $Q$, the Hellinger distance between $P$ and $P'$ is:

$$d_H(P, P') = \sqrt{\frac{1}{2}\int_\Omega \left(\sqrt{\frac{dP}{dQ}} - \sqrt{\frac{dP'}{dQ}}\right)^2 dQ}$$

Let $\psi(p) = \sum_{i=0}^{k-1} \mathcal{M}(p_i)\Delta M_i - \frac{1}{2}\sum_{i=0}^{k-1} \mathcal{M}^2(p_i)\Delta\tau$. Then

$$|\psi(p) - \psi(p')| = \left|\sum_{i=0}^{k-1} \mathcal{M}(p_i)\Delta M_i - \mathcal{M}(p'_i)\Delta M'_i - \frac{1}{2}\left(\sum_{i=0}^{k-1} \mathcal{M}^2(p_i)\Delta\tau - \mathcal{M}^2(p'_i)\Delta\tau\right)\right|$$

$$\leq 2\|\mathcal{M}\|\sum_{i=0}^{k-1} \Delta M_i^{\max} + \|\mathcal{M}\|^2 k\Delta\tau$$

where $\Delta M_i^{\max} = \max(|\Delta M_i|, |\Delta M_i'|)$. In the last step, we have made use of the fact that $\mathcal{M}: \mathbb{R} \to \mathbb{R}$ being the coefficient of an SDE is a bounded and Lipschitz continuous function in $p$ and $\|.\|$ is the essential supremum norm given by

$$\|\mathcal{M}\| = \underset{p \in \mathbb{R}^{n_p}}{\text{ess sup}} |\mathcal{M}(p)|$$

defined on the set of all real-valued continuous bounded functions, $C_b(\mathbb{R})$.

*Theorem 1*

Let $k \in \mathbb{N}$ be the total number of iterations and $\Delta M_i^{\max} = \max(|\Delta M_i|, |\Delta M_i'|)$, $\Delta M_i^{\max} > 0, i = 0, ..., k-1$. Then there exists $C = C(k, \|\mathcal{M}\|, \Delta \tau)$, $C > 0$ such that $d_H(\mathrm{P}, \mathrm{P}') \leq C \exp\left(\dfrac{\|\mathcal{M}\|}{2} \sum_{i=0}^{k-1} \Delta M_i^{\max}\right) \|\mathcal{M}\| \sqrt{k \Delta \tau}$.

Furthermore, if $f: \mathbb{R} \to \mathbb{R}$ is a measurable function such that $f \in L_\mathrm{P}^2(\Omega) \cap L_{\mathrm{P}'}^2(\Omega)$, then there exists $C' = C'(|f|, k, \|\mathcal{M}\|, \Delta \tau)$, $C' > 0$ such that

$$\left| E_\mathrm{P}[f] - E_{\mathrm{P}'}[f] \right| \leq C' \exp\left(\frac{\|\mathcal{M}\|}{2} \sum_{i=0}^{k-1} \Delta M_i^{\max}\right) \|\mathcal{M}\| \sqrt{k \Delta \tau}.$$

Proof:

$$\begin{aligned}
\left(d_H(\mathrm{P}, \mathrm{P}')\right)^2 &= \frac{1}{2} \int_\Omega \left(\sqrt{\Lambda_k} - \sqrt{\Lambda_k'}\right)^2 dQ \\
&= \frac{1}{2} \int_\Omega \left(\exp\left(\frac{1}{2}\psi(p)\right) - \exp\left(\frac{1}{2}\psi(p')\right)\right)^2 dQ \\
&\leq \frac{1}{8} \int_\Omega \exp\left(\|\mathcal{M}\| \sum_{i=0}^{k-1} \Delta M_i^{\max}\right) |\psi(p) - \psi(p')|^2 dQ
\end{aligned}$$

The last inequality is arrived at using the local Lipschitz property of the exponential function and the bound for $\psi(p) \leq \|\mathcal{M}\| \sum_{i=1}^{k} \Delta M_i^{\max}$. Now, using the inequality above, we get,

$$\left(d_H(P,P')\right)^2 \leq \frac{1}{8}\exp\left(\|\mathcal{M}\|\sum_{i=0}^{k-1}\Delta M_i^{\max}\right)\int_\Omega \left(2\|\mathcal{M}\|\sum_{i=0}^{k-1}\Delta M_i^{\max} + \|\mathcal{M}\|^2 k\Delta\tau\right)^2 dQ$$

$$\leq \frac{1}{4}\exp\left(\|\mathcal{M}\|\sum_{i=0}^{k-1}\Delta M_i^{\max}\right)\left\{\int_\Omega \left(2\|\mathcal{M}\|\sum_{i=0}^{k-1}\Delta M_i^{\max}\right)^2 dQ + \int_\Omega \left(\|\mathcal{M}\|^2 k\Delta\tau\right)^2 dQ\right\}$$

$$\leq \frac{1}{4}\exp\left(\|\mathcal{M}\|\sum_{i=0}^{k-1}\Delta M_i^{\max}\right)\left\{4\|\mathcal{M}\|^2 E_Q\left[\left(\sum_{i=0}^{k-1}\Delta M_i^{\max}\right)^2\right] + \|\mathcal{M}\|^4 k^2(\Delta\tau)^2\right\}$$

$$\leq 2^{k-3}\exp\left(\|\mathcal{M}\|\sum_{i=0}^{k-1}\Delta M_i^{\max}\right)\left\{4\|\mathcal{M}\|^2 \sum_{i=0}^{k-1} E_Q\left[\left(\Delta M_i^{\max}\right)^2\right] + \|\mathcal{M}\|^4 k^2(\Delta\tau)^2\right\}$$

$$\leq 2^{k-3}\exp\left(\|\mathcal{M}\|\sum_{i=0}^{k-1}\Delta M_i^{\max}\right)\left\{4\|\mathcal{M}\|^2 k\Delta\tau + \|\mathcal{M}\|^4 k^2(\Delta\tau)^2\right\}$$

$$\leq C_1 \exp\left(\|\mathcal{M}\|\sum_{i=0}^{k-1}\Delta M_i^{\max}\right)\|\mathcal{M}\|^2 k\Delta\tau$$

where $C_1 = 2^{k-3}\left(4 + \|\mathcal{M}\|^2 k\Delta\tau\right)$.

Hence, we get
$$d_H(P,P') \leq C\exp\left(\frac{\|\mathcal{M}\|}{2}\sum_{i=0}^{k-1}\Delta M_i^{\max}\right)\|\mathcal{M}\|\sqrt{k\Delta\tau}. \qquad (25)$$

where $C = \sqrt{C_1}$. If $f \in L^2_P(\Omega) \cap L^2_{P'}(\Omega)$, then the bounds in terms of the Hellinger metric could be converted in terms of expectations [16] as

$$\left|E_P[f] - E_{P'}[f]\right| \leq 2\left(E_P\left[|f|^2\right] + E_{P'}\left[|f|^2\right]\right)^{\frac{1}{2}} d_H(P,P')$$

which gives

$$\left|E_P[f] - E_{P'}[f]\right| \leq 2C\left(E_P\left[|f|^2\right] + E_{P'}\left[|f|^2\right]\right)^{\frac{1}{2}} \exp\left(\frac{\|\mathcal{M}\|}{2}\sum_{i=0}^{k-1}\Delta M_i^{\max}\right)\|\mathcal{M}\|\sqrt{k\Delta\tau}$$

This completes the proof of Theorem 1.

**Remark 1**: The bound in Eqn. (25) is dependent on $\Delta\tau$ that arises as a result of the variance of the Brownian motion generating the fictitious measurements. As a result, variations in the estimates owing to small perturbations in the measurement can be controlled using $\Delta\tau$. In other words, the extra noise that is added to the static measurements acts as a regularizer to stabilize the solutions. Indeed, the proposed method is general enough to accommodate non-Brownian and non-Gaussian variants of the regularizing noise processes. For instance, one could use Poisson martingale representation of the measurement-prediction mismatch to arrive at novel (and possibly even more efficient) variants of the current scheme.

**3.2 Convergence of the particle-approximated and $\tau$ - discretized KS equation**

The aim here is to obtain a convergence result for the realization-wise (or particle-wise) approximated and $\tau$-discretized KS equation. As was mentioned in the beginning of this section, the order of convergence is arrived at by separating the two approximations, the one in $\tau$ and the other in the finiteness of the ensemble. The combined result is obtained by summing the two error bounds, one of which corresponds to the convergence of the $\tau$-discretized scheme to the $\tau$-continuous model. The second error bound describes the convergence of the particle approximation to the $\tau$-discretized exact law of the parameter.

In [17, 18], the order of convergence of Euler-type time-discretization schemes to the continuous time filtering model has been studied. The evolutionary global search method proposed in this work involves the evolution of the conditioned process over a continuous parameter $\tau$ that is analogous to time in filtering. The idea behind the proof provided in [18] has therefore been made use of to arrive at an error bound for the approximation in $\tau$.

The convergence of particle-based empirical approximations to exact probability densities have been studied extensively in [19, 20] by establishing central limit theorems for the same. Most of these studies are based on the Bayesian approach of particle filtering involving the prior, the weighted posterior and sometimes resampling. In [21], the authors have estimated the order of convergence of the particle approximation of the Kushner-Stratonovich equation in nonlinear filtering. However, they have employed the conventional weight-based update strategy in contrast to the additive scheme proposed here. Here the order of convergence of the particle approximated conditioned process, updated via the KSG as in Eqn. (20), is studied. For a lucid presentation, the proof below is for the case of $n_p = 1$ and $n_M = 1$.

*Theorem 2*

Let $f : \mathbb{R} \to \mathbb{R}$ be a bounded and twice continuously differentiable measurable function of $p$, i.e. $f \in C_b^2(\mathbb{R})$. Then for almost every trajectory $\omega$ and any $\varepsilon > 0$ there exists $C(p, \omega) > 0$ such that

$$\left|\hat{\pi}_k^d(f) - \pi_k(f)\right| \leq \frac{k\|f\|}{\sqrt{n_E}} + (k-1)(k-2)\Delta\tau C_f + 4\|\mathcal{M}\|\|f\|\sum_{i=1}^{k}\sum_{j=k-i+2}^{k}\left|\Delta M_j - \hat{\pi}_j^d(\mathcal{M}(p))\Delta\tau\right|$$

$$+ C(p,\omega)(\Delta\tau)^{\frac{1}{2}-\varepsilon}$$

where $\hat{\pi}_k^d := \hat{\pi}_{\tau_k}^d$ is the ensemble mean operator introduced in Section 2 and $\pi_k$ the exact law at $\tau_k$. Also $C_f = \frac{\Sigma_B^2}{2}\sup_{p\in\mathbb{R}^{n_p}}\left|\frac{d^2 f}{dp^2}\right|$ and $C(p,\omega)$ does not depend on $\Delta\tau$.

Proof:

We can separate the two approximations as follows:

$$\left|\hat{\pi}_k^d(f) - \pi_k(f)\right| \leq \left|\hat{\pi}_k^d(f) - \pi_k^d(f)\right| + \left|\pi_k^d(f) - \pi_k(f)\right|$$

where $\pi_k^d(f)$ is the $\tau$-discretized exact law of $f$. We first consider the particle approximation to the $\tau$-discretized conditioned estimate.

$$|\hat{\pi}_k^d(f) - \pi_k^d(f)| = |\hat{\pi}_k^d(f) - \left\{\hat{\pi}_{k-1}^d(f) + \int_{\tau_{k-1}}^{\tau_k}\pi_s(Lf)ds + \int_{\tau_{k-1}}^{\tau_k}G_s^{KS}(dM_s - \pi_s(\mathcal{M}(p))ds)\right\}$$

$$+ \left\{\hat{\pi}_{k-1}^d(f) + \int_{\tau_{k-1}}^{\tau_k}\pi_s(Lf)ds + \int_{\tau_{k-1}}^{\tau_k}G_s^{KS}(dM_s - \pi_s(\mathcal{M}(p))ds)\right\}$$

$$- \left\{\hat{\pi}_{k-2}^d(f) + \int_{\tau_{k-2}}^{\tau_k}\pi_s(Lf)ds + \int_{\tau_{k-2}}^{\tau_k}G_s^{KS}(dM_s - \pi_s(\mathcal{M}(p))ds)\right\}$$

$$+ \{\ldots\} - \{\ldots\}$$

$$+ \ldots$$

$$+ \left\{\hat{\pi}_1^d(f) + \int_{\tau_1}^{\tau_k}\pi_s(Lf)ds + \int_{\tau_1}^{\tau_k}G_s^{KS}(dM_s - \pi_s(\mathcal{M}(p))ds)\right\}$$

$$- \left\{\pi_0(f) + \int_{\tau_0}^{\tau_k}\pi_s(Lf)ds + \int_{\tau_0}^{\tau_k}G_s^{KS}(dM_s - \pi_s(\mathcal{M}(p))ds)\right\}|$$

where $\{\tau_0, \tau_1, \ldots, \tau_k\}$ is a partition of $[0, \tau_k]$ such that $0 = \tau_0 < \tau_1 < \ldots < \tau_{k-1} < \tau_k$, $\tau_i - \tau_{i-1} = \Delta\tau$, $i = 1, 2, \ldots, k$, $L_s f = \frac{\Sigma_B^2}{2}\frac{d^2 f}{dp_s^2}$ and $G_s^{KS} = (\pi_s(\mathcal{M}(p)f) - \pi_s(\mathcal{M}(p))\pi_s(f))$.

Also, $\pi_0(f)$ is the initial estimate of $f$ given by $\pi_0(f) = E_P[f(p_0) | M_0 = 0]$, which is assumed to be known. Given that $\pi_k^d$ is not known *a-priori*, the above decomposition on the LHS uses the empirically computed law $\hat{\pi}_k^d$ in a telescoping sum in order to arrive at $\pi_k^d$. In the absence of numerical approximations to the integrals involved in the sum, however, we end up with the term

$$\left\{ \pi_0(f) + \int_{\tau_0}^{\tau_k} \pi_s(Lf) ds + \int_{\tau_0}^{\tau_k} G_s^{KS}\left(dM_s - \pi_s(\mathcal{M}(p))ds\right) \right\}$$

which is identical with $\pi_k(f)$. Nevertheless, upon Euler-approximating the above integrals, we obtain

$$|\hat{\pi}_k^d(f) - \pi_k^d(f)| = |\hat{\pi}_k^d(f) - \left\{\hat{\pi}_{k-1}^d(f) + L_{k-1}f\Delta\tau + G_k^{KS}\left(\Delta M_k - \hat{\pi}_k^d(\mathcal{M}(p))\Delta\tau\right)\right\}$$
$$+ \left\{\hat{\pi}_{k-1}^d(f) + L_{k-1}f\Delta\tau + G_k^{KS}\left(\Delta M_k - \hat{\pi}_k^d(\mathcal{M}(p))\Delta\tau\right)\right\}$$
$$- \left\{\begin{array}{l} \hat{\pi}_{k-2}^d(f) + L_{k-2}f\Delta\tau + G_{k-1}^{KS}\left(\Delta M_{k-1} - \hat{\pi}_{k-1}^d(\mathcal{M}(p))\Delta\tau\right) \\ + L_{k-1}f\Delta\tau + G_k^{KS}\left(\Delta M_k - \hat{\pi}_k^d(\mathcal{M}(p))\Delta\tau\right) \end{array}\right\}$$
$$+ \{...\} - \{...\}$$
$$+ ...$$
$$+ \left\{\begin{array}{l} \hat{\pi}_1^d(f) + L_1 f\Delta\tau + G_2^{KS}\left(\Delta M_2 - \hat{\pi}_2^d(\mathcal{M}(p))\Delta\tau\right) \\ \quad + L_2 f\Delta\tau + G_3^{KS}\left(\Delta M_3 - \hat{\pi}_3^d(\mathcal{M}(p))\Delta\tau\right) \\ \quad + ... \\ \quad + L_{k-1}f\Delta\tau + G_k^{KS}\left(\Delta M_k - \hat{\pi}_k^d(\mathcal{M}(p))\Delta\tau\right) \end{array}\right\}$$
$$- \left\{\begin{array}{l} \pi_0(f) + L_0 f\Delta\tau + G_1^{KS}\left(\Delta M_1 - \hat{\pi}_1^d(\mathcal{M}(p))\Delta\tau\right) \\ \quad + L_1 f\Delta\tau + G_2^{KS}\left(\Delta M_2 - \hat{\pi}_2^d(\mathcal{M}(p))\Delta\tau\right) \\ \quad + ... \\ \quad + L_{k-1}f\Delta\tau + G_k^{KS}\left(\Delta M_k - \hat{\pi}_k^d(\mathcal{M}(p))\Delta\tau\right) \end{array}\right\} |$$

Here $L_{i-1}f = \dfrac{\Sigma_B^2}{2}\dfrac{d^2 f}{d\hat{p}_{i-1}^2}$ and $G_i^{KS} = \left(\hat{\pi}_i^d(\mathcal{M}(p)f) - \hat{\pi}_i^d(\mathcal{M}(p))\hat{\pi}_i^d(f)\right)$ for $i = 1,2,...,k$. Note that in an expression such as $\left\{\hat{\pi}_{k-1}^d(f) + L_{k-1}f\Delta\tau + G_k^{KS}\left(\Delta M_k - \hat{\pi}_k^d(\mathcal{M}(p))\Delta\tau\right)\right\}$, while $\hat{\pi}_{k-1}^{d}(.)$ operates on the updated particles at $\tau_{k-1}$, $\hat{\pi}_{k}^{d}(.)$ operates only on the predicted particles at $\tau_{k}$. Let $T_i$ denote the $i^{th}$ term in the last expression for $|\hat{\pi}_{k}^{d}(f) - \pi_{k}^{d}(f)|$, i.e.

$$T_i = \begin{Bmatrix} \hat{\pi}_{k-i+1}^{d}(f) + L_{k-i+1}f\Delta\tau + G_{k-i+2}\left(\Delta M_{k-i+2} - \hat{\pi}_{k-i+2}^{d}(\mathcal{M}(p))\Delta\tau\right) \\ + ... \\ + L_{k-1}f\Delta\tau + G_{k}\left(\Delta M_{k} - \hat{\pi}_{k}^{d}(\mathcal{M}(p))\Delta\tau\right) \end{Bmatrix} - \begin{Bmatrix} \hat{\pi}_{k-i}^{d}(f) + L_{k-i}f\Delta\tau + G_{k-i+1}\left(\Delta M_{k-i+1} - \hat{\pi}_{k-i+1}^{d}(\mathcal{M}(p))\Delta\tau\right) \\ + ... \\ + L_{k-1}f\Delta\tau + G_{k}\left(\Delta M_{k} - \hat{\pi}_{k}^{d}(\mathcal{M}(p))\Delta\tau\right) \end{Bmatrix}$$

The expression for $T_i$ may be rewritten as

$$T_i = \hat{\pi}_{k-i+1}^{d}(f) + \sum_{j=k-i+2}^{k} L_{j-1}f\Delta\tau + G_j\left(\Delta M_j - \hat{\pi}_j^{d}(\mathcal{M}(p))\Delta\tau\right) - \left\{\hat{\pi}_{k-i}^{d}(f) + \sum_{j=k-i+1}^{k} L_{j-1}f\Delta\tau + G_j\left(\Delta M_j - \hat{\pi}_j^{d}(\mathcal{M}(p))\Delta\tau\right)\right\}$$

i.e.

$$T_i = \hat{\pi}_{k-i+1}^{d}(f) + \sum_{j=k-i+2}^{k} L_{j-1}f\Delta\tau + G_j\left(\Delta M_j - \hat{\pi}_j^{d}(\mathcal{M}(p))\Delta\tau\right) - \left\{\pi_{k-i+1}^{d}(f) + \sum_{j=k-i+2}^{k} L_{j-1}f\Delta\tau + G_j\left(\Delta M_j - \hat{\pi}_j^{d}(\mathcal{M}(p))\Delta\tau\right)\right\}$$

Let $S_j(f, y)$ be defined as

$$S_j(f, y) = L_{j-1}f\Delta\tau + G_j\left(\Delta M_j - \hat{\pi}_j^{d}(\mathcal{M}(p))\Delta\tau\right)$$

so that

$$|S_j(f, y)| \leq \left(C_f \Delta\tau + 2\|\mathcal{M}\|\|f\|\left|\Delta M_j - \hat{\pi}_j^{d}(\mathcal{M}(p))\Delta\tau\right|\right)$$

where $C_f = \frac{\Sigma_B^2}{2} \sup_{p \in \mathbb{R}^{n_p}} \left|\frac{d^2 f}{dp^2}\right|$ and the bound for the update is obtained using the expression $G_j^{KS} = \left(\hat{\pi}_j^d(\mathcal{M}(p)f) - \hat{\pi}_j^d(\mathcal{M}(p))\hat{\pi}_j^d(f)\right)$. Then, by the triangle inequality,

$$|T_i| \leq \left|\hat{\pi}_{k-i+1}^d(f) - \pi_{k-i+1}^d(f)\right| + 2\sum_{j=k-i+2}^{k} |S_j(f,y)|$$

i.e. $|T_i| \leq \left|\hat{\pi}_{k-i+1}^d(f) - \pi_{k-i+1}^d(f)\right| + 2\sum_{j=k-i+2}^{k} \left(C_f \Delta\tau + 2\|\mathcal{M}\|\|f\|\left|\Delta M_j - \hat{\pi}_j^d(\mathcal{M}(p))\Delta\tau\right|\right)$

Now, using the fact that $\hat{\pi}_{k-i+1}^d$ is the ensemble mean of $n_E$ independent random variables whose (exact) conditioned estimate is $\pi_{k-i+1}^d$, we have $\left|\hat{\pi}_{k-i+1}^d(f) - \pi_{k-i+1}^d(f)\right| \leq \frac{\|f\|}{\sqrt{n_E}}$. Therefore, we get

$$|T_i| \leq \frac{\|f\|}{\sqrt{n_E}} + 2\Delta\tau(i-2)C_f + 4\|\mathcal{M}\|\|f\|\sum_{j=k-i+2}^{k}\left|\Delta M_j - \hat{\pi}_j^d(\mathcal{M}(p))\Delta\tau\right|$$

By summing the terms and using $\left|\hat{\pi}_k^d(f) - \pi_k^d(f)\right| \leq \sum_{i=1}^{k}|T_i|$, we get

$$\left|\hat{\pi}_k^d(f) - \pi_k^d(f)\right| \leq \frac{k\|f\|}{\sqrt{n_E}} + (k-1)(k-2)\Delta\tau C_f + 4\|\mathcal{M}\|\|f\|\sum_{i=1}^{k}\sum_{j=k-i+2}^{k}\left|\Delta M_j - \hat{\pi}_j^d(\mathcal{M}(p))\Delta\tau\right|$$

(26)

Now, we proceed to obtain the convergence result for the $\tau$-discretization. As mentioned earlier, the proof is on the same lines as Theorem 3.1 of [18]. It has been proved in [17] that for $q \geq 1$ and for some $C > 0$

$$\left(E\left[\left|p_k^d - p_k\right|^{2q}\right]\right)^{\frac{1}{2q}} = C\sqrt{\Delta\tau}$$

where $p_k^d$ is the $\tau$-discretized version of the parameter process $p_k$.

$$E_{\mathrm{P}}\left[\left|\pi_k^d(f) - \pi_k(f)\right|^{2q}\right] = E_{\mathrm{P}}\left[\left|E_{\mathrm{P}}\left[f(p_k^d) \mid \mathcal{F}_k\right] - E_{\mathrm{P}}\left[f(p_k) \mid \mathcal{F}_k\right]\right|^{2q}\right]$$

$$\leq E_{\mathrm{P}}\left[E_{\mathrm{P}}\left[\left|f(p_k^d) - f(p_k)\right| \mid \mathcal{F}_k\right]\right]^{2q}$$

$$\leq E_{\mathrm{P}}\left[E_{\mathrm{P}}\left[\left|f(p_k^d) - f(p_k)\right|^{2q} \mid \mathcal{F}_k\right]\right]$$

$$\leq E_{\mathrm{P}}\left[\left|f(p_k^d) - f(p_k)\right|^{2q}\right]$$

The first two inequalities above are obtained via the conditional version of Jensen's inequality and the last one via the smoothing property of conditional expectation. Now, using the fact that $f \in C_b^2(\mathbb{R})$ and assuming that the derivatives of $f$ satisfy the polynomial growth condition, we have [18]

$$\left|f(p_k^d) - f(p_k)\right| \leq K\left(1 + |p_k^d|^a + |p_k|^a\right)|p_k^d - p_k|$$

where $K$ and $a$ are positive constants. Now, since $p_k^d$ and $p_k$ have bounded moments of any order, we obtain,

$$E_{\mathrm{P}}\left[\left|f(p_k^d) - f(p_k)\right|^{2q}\right] \leq K E_{\mathrm{P}}\left[\left(1 + |p_k^d|^a + |p_k|^a\right)^{2q}|p_k^d - p_k|^{2q}\right]$$

Applying Cauchy-Schwarz inequality, we get

$$E_{\mathrm{P}}\left[\left|f(p_k^d) - f(p_k)\right|^{2q}\right] \leq K\sqrt{E_{\mathrm{P}}\left[\left(1 + |p_k^d|^a + |p_k|^a\right)^{4q}\right]}\sqrt{E_{\mathrm{P}}\left[|p_k^d - p_k|^{4q}\right]}$$

$$\leq C'(p)(\Delta \tau)^q$$

i.e.

$$E_{\mathrm{P}}\left[\left|\pi_k^d(f) - \pi_k(f)\right|^{2q}\right]^{\frac{1}{2q}} \leq C'(p)(\Delta \tau)^{\frac{1}{2}}$$

Use of the Markov inequality and the above expression yields

$$\mathrm{P}\left(\left|\pi_k^d(f) - \pi_k(f)\right| > (\Delta \tau)^\gamma\right) \leq \frac{E_{\mathrm{P}}\left[\left|\pi_k^d(f) - \pi_k(f)\right|^{2q}\right]}{(\Delta \tau)^{2q\gamma}}$$

$$\leq C'(p)(\Delta \tau)^{q(1-2\gamma)}$$

Then for any $\gamma = \frac{1}{2} - \varepsilon$, we have

$$\sum_{k=1}^{\infty} P\left(\left|\pi_k^d(f) - \pi_k(f)\right| > \left(\frac{\tau_k}{k}\right)^{\gamma}\right) \leq C'(p)(\tau_k)^{p(1-2\gamma)} \sum_{k=1}^{\infty} \frac{1}{k^{p(1-2\gamma)}} < \infty$$

by choosing a sufficiently large $q$. Hence, due to the Borel-Cantelli Lemma, the random variable $\varsigma := \sup_{\Delta\tau > 0} (\Delta\tau)^{-\gamma} \left|\pi_k^d(f) - \pi_k(f)\right|$ is finite almost surely. Finally we have the following result:

$$\left|\pi_k^d(f) - \pi_k(f)\right| \leq C(p,\omega)(\Delta\tau)^{\frac{1}{2}-\varepsilon}$$

This completes the proof of Theorem 2.

## 4. Numerical Illustrations: The Inverse Problem of UMOT

A brief account of the inverse problem of UMOT has been provided in Appendix I. We have used a two-dimensional object of circular cross-section of radius 4cm in all our numerical simulations. The background optical and mechanical properties are taken as $\mu_a = 0.1 \text{cm}^{-1}$, $\mu_s' = 8$ cm$^{-1}$ and $D_B = 10^{-9}$ cm$^2$sec. A cross-section of a hyperboloid region centered at (0, 0) cm (which is the centre of the object), of waist radius 0.1 cm and height 1.5 cm is set as the insonified region $IR$, where $p$ has a value of $1 \times 10^{-7}$ cm$^2$. $p$ is assumed to be zero outside the ROI. The ROI has two cases of inhomogeneous inclusions: one at its centre and the other at its periphery. For the centrally located inhomogeneity, $p$ is $3 \times 10^{-7}$ cm$^2$ and for those at the periphery the values are $2 \times 10^{-7}$ cm$^2$ (at the top) and $3 \times 10^{-7}$ cm$^2$ (at the bottom). The ultrasound frequency is set at 1MHz. For simulating 'experimental' data, we solve the forward equation (Eqns. (A6) and (A7)) after finite element (FE) discretization with 2773 nodes and 5376 elements to cover the object. The forward equation is solved for $\vartheta$ ranging from 0s to $5 \times 10^{-7}$s with an increment of $1.25 \times 10^{-7}$s to generate 4 samples. The measurements $\{G^\delta(\mathbf{r}, \vartheta)\}$ are collected with 21 equi-angularly placed detectors covering a total angular span of 180 degrees with the central detector placed diametrically opposite to the source. The source-detector combination is rotated in unison by steps of $30^0$ and 12 sets of 21 detector readings are taken in order to generate a complete data set. Our measurement on $G^\delta(\mathbf{r},\vartheta)$ is obtained by time Fourier transforming $G^\delta(\mathbf{r},\vartheta)$ and taking the modulus of the Fourier transform at $\varpi = \varpi_a$, the frequency of the ultrasound. To this numerically computed data, we have added 1% Gaussian noise to generate a set of 'measurements'.

The reconstruction algorithm is initiated with $p$ at its background value of $1 \times 10^{-7}$ cm$^2$. For reconstruction through the GN approach, the regularization parameter $\beta$ is fixed as the maximum of the diagonal of ($J^T J$) initially, and then is adaptively reduced by a factor of 2 at each iteration if the error $\chi_{k+1}$ is less than $\chi_k$. The algorithm is terminated when

$$100 \times \left( \frac{\chi_{k+1} - \chi_k}{\chi_k} \right) < 0.1$$

The original distribution of $p(\mathbf{r})$ used to generate the experimental data for the side and central inhomogeneities are shown in Figures 1(a) and 1(b). The proposed method has been implemented using both KSG and LSG as per Version 2 discussed in Section 2. The recovered $p(\mathbf{r})$ profiles (plotted as the first moments of the associated conditional probability densities) along with cross-sections through the centre of the inhomogeneity, corresponding to the ROI with inhomogeneities at the side, are shown in Fig. 2. Whereas the reconstructions are in Figs. 2(a), 2(b) and 2(d) (respectively for GN, KSG and LSG) the corresponding cross-sectional plots are in Figs. 2(c) and 2(e). Similar results corresponding to the object with the central inhomogeneity are shown in Figure 3. The proposed method clearly gives a better reconstruction in comparison to the GN algorithm especially for the central inhomogeneity, wherein, as mentioned earlier, the sensitivity of the measurements to changes in $p$ is poor. Note that we have used $\alpha_1 = 2$ uniformly for KSG and $\alpha_1 = 3$ (side inhomogeneity) and $\alpha_1 = 5$ (central inhomogeneity) for LSG to begin the iterations and successively reduced it to zero following the schedule mentioned in Section 3. Also, due to the large diffusions given to the parameters initially, a few of the realizations could be in the form of outliers that tend to amplify the measurement-prediction misfit and precipitate possible divergence of the algorithm over subsequent iterations. Hence, we have employed a somewhat unconventional rejection strategy (RS), namely, each particle at the current iteration is updated according to Eqns. (22) and (23) only if the corresponding realization of the objective functional (i.e. the sum of squares of the individual misfit components, as in Version 1) is lower than that at the immediately preceding iteration. Otherwise, no updating is applied to the particles, which are passed on to the prediction step of the next iteration. The RS has been employed only with LSG for the case of central inhomogeneity and a comparison of results with and without RS is shown in Figure (3e).

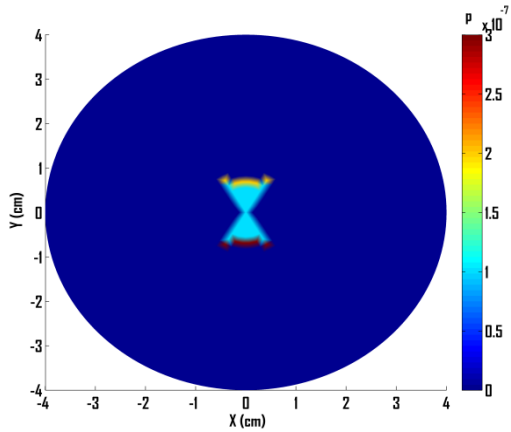 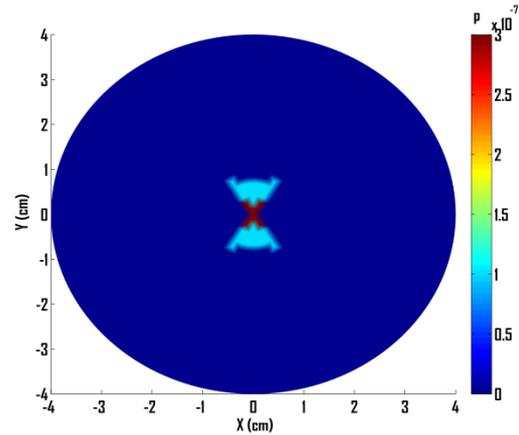

Figure 1(a): Grey level image of the object with side inhomogeneity in terms of $p(\mathbf{r})$

Figure 1(b): Grey level image of the object with central inhomogeneity in terms of $p(\mathbf{r})$

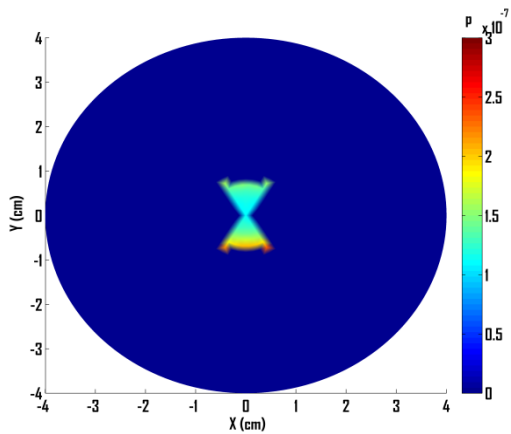 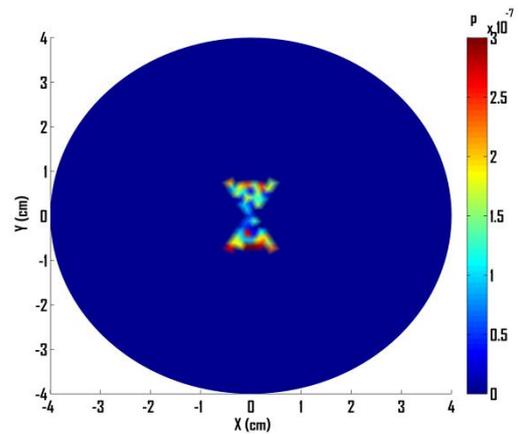

Figure 2(a): Reconstructed image corresponding to the object in Fig. 1(a) obtained from the GN algorithm

Figure 2(b): Reconstructed image corresponding to the object in Fig. 1(a) obtained using KSG

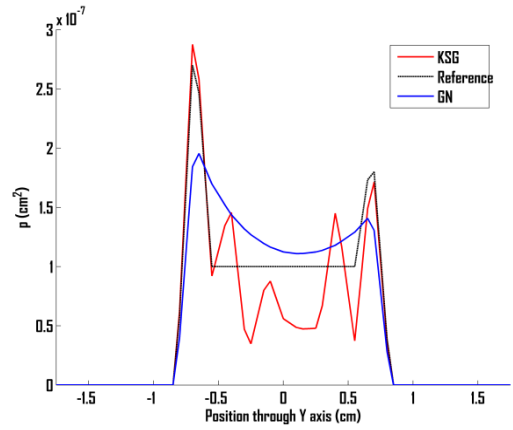

Figure 2(c): Cross-sectional plot through the center of the inhomogeneity in Fig. 1(a)

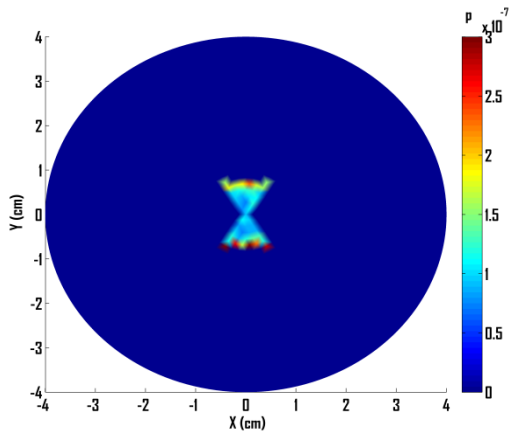

Figure 2(d): Reconstructed image corresponding to the object in Fig. 1(a) obtained using LSG

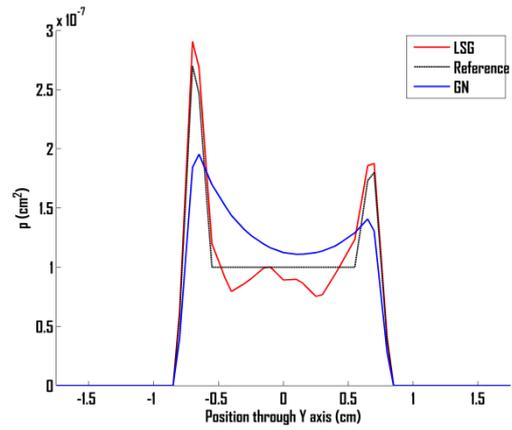

Figure 2(e): Cross-sectional plot through the center of the inhomogeneity in Fig. 1(a)

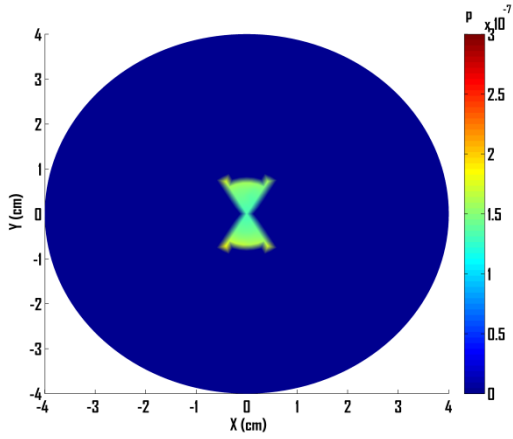 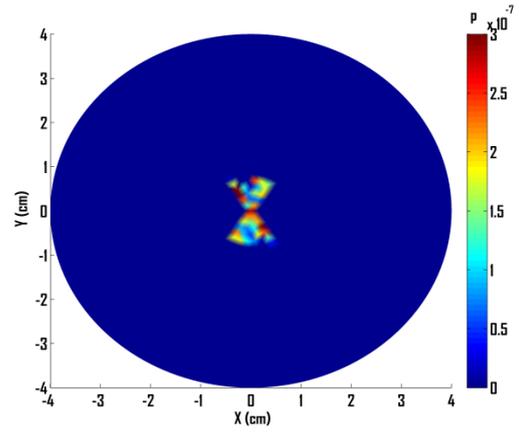

Figure 3(a): Reconstructed image corresponding to the object in Fig. 1(b) obtained using GN

Figure 3(b): Reconstructed image corresponding to the object in Fig. 1(b) obtained using KSG

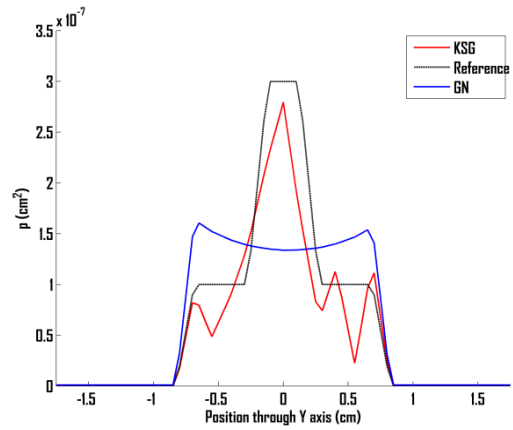

Figure 3(c): Cross-sectional plot through the center of the inhomogeneity in Fig. 1(b)

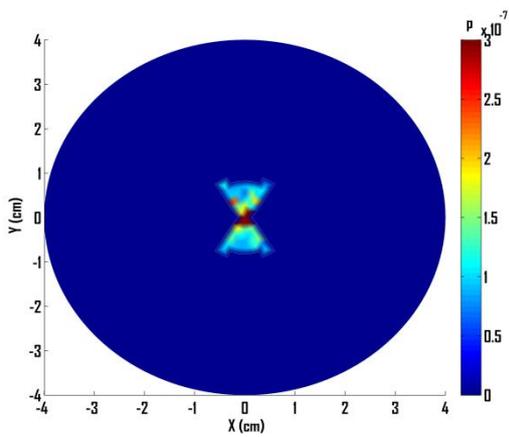 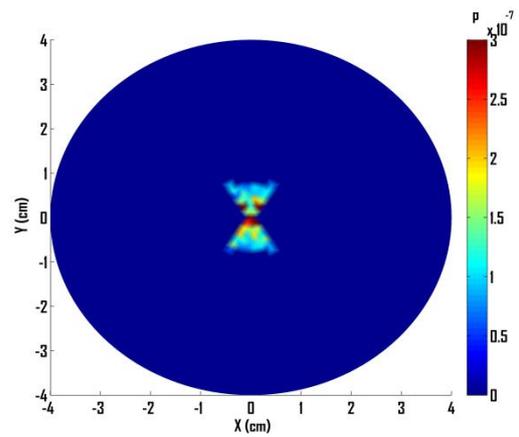

Figure 3(d): Reconstructed image corresponding to the object in Fig. 1(b)

Figure 3(e): Reconstructed image corresponding to the object in Fig. 1(b)

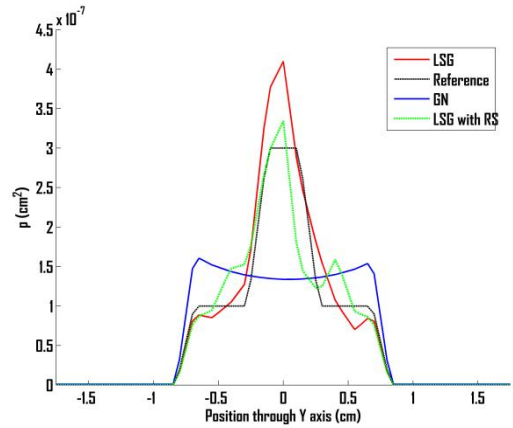

Figure 3(f): Cross-sectional plot through the center of the inhomogeneity in Fig. 1(b)

In the past, quantitative recovery of parameters, in the context of UMOT, meant only an average value corresponding to the US focal volume. To the best of our knowledge, it was in [8] that a space-resolved recovery of a mechanical property was first attempted. The regularized GN scheme conspicuously showed the limitations of a (semi-)deterministic scheme in handling a severely ill-posed reconstruction from noisy and sparse data. The evolutionary Bayesian search method introduced here is seen to be the answer to a fairly broad class of ill-posed reconstruction problems of which UMOT is just an example. The recovery of parameter in the central region where the sensitivity of measurement to parameter variations is poor is a pointer to what a well-conceived stochastic search scheme can do when most deterministic methods fail. The contrast recovery from all the variants of the stochastic search scheme is excellent and the background noise within limits.

Though not explicitly reported above, it may be worthwhile to mention that a few other competing stochastic methods, e.g. the GA or the Markov Chain Monte Carlo (MCMC), typically take far more iterations for convergence vis-a-vis the present method. The vastly superior computational performance with our current proposal may be traced to the explicitly built-in directional information in the nonlinear update terms.

**5. Concluding Remarks**

The generalized Bayesian search method, proposed here for the solution of inverse problems, falls in the broad category of evolutionary random search techniques, e.g. the genetic algorithm. Compared with the latter, however, the proposed method crucially differs in bringing in a new class of stochastic characterizations towards extremizing the cost functional (or the error functions, as appropriate), which together with the powerful notion of a change of measure and the machinery of stochastic calculus yields an additive update term to recursively direct and hence accelerate the global search procedure. Remarkably enough, the directional information so obtained is strictly derivative-free (non-Newtonian) and is borne out of the statistical moment information contained in the artificially introduced diffusive character of the parameters to be reconstructed. The global search may be further aided by tuning the so-called intensity of the additive correction term through an annealing-type scalar parameter. An interesting aspect of the convergence/stability analysis undertaken here is the fact that the reconstructed parameter profile can be made practically insensitive to small variations in the acquired data by keeping the intensity of the regularizing noise process sufficiently low, a regime wherein the proposed schemes are found to remain numerically robust. This may be contrasted with traditional weight-based Bayesian search approaches, wherein an inadequate dispersion of the weights owing to the low intensity of the regularizing noise would precipitate the problem of particle degeneracy. In the specific context of the UMOT problem considered here, the contrast in the performance of the proposed method vis-à-vis the regularized Gauss-Newton method is particularly highlighted when the inhomogeneous inclusion is positioned around the centre of the ROI, a case characterized by a low sensitivity of the conventional Gateaux derivative with respect to variations in the parameter profile.

The current proposal on the evolutionary Bayesian search scheme may be extended, perhaps non-trivially, by characterizing the measurement-prediction error using non-Brownian martingales. Prominent amongst such possibilities is the Poisson martingale, which, unlike a Brownian martingale, has the added advantage of having zero quadratic variation (i.e. with sample paths of finite variation) that in turn enables strictly bounded and controlled realizations of the regularizing noise variables. Such a possibility is currently under exploration.

# APPENDIX I: A Background of the UMOT

The UMOT [22 23] has been introduced as a remedy for poor spatial resolution in the optical contrast recovery available from diffuse optical tomography. Here a tightly focused ultrasound (US) beam introduces a modulation of the refractive index $(n(\mathbf{r}))$ and the mean position of the scattering centres in a localized region referred to as the region of interest (ROI) in the object to be imaged [24]. A coherent light beam interrogating the object picks up a phase modulation from the insonified ROI, which modulates the overall decay of the specific intensity, $I(\mathbf{r}, \hat{k}_s, \vartheta)$. ($I(\mathbf{r}, \hat{k}_s, \vartheta)$ is derived from the mutual coherence function of light, $<E_a(\mathbf{r}_a,t)E_b^*(\mathbf{r}_b,t+\vartheta)>$ expressed in centre of gravity co-ordinates (i.e., $\mathbf{r} = (\mathbf{r}_b + \mathbf{r}_a)/2$) with $\hat{k}_s$ representing the normalized scattered light propagation vector in the direction $\mathbf{r}_b - \mathbf{r}_a$ [25].) The property of light we follow here is an angle-averaged version of $I(\mathbf{r}, \hat{k}_s, \vartheta)$, called the amplitude autocorrelation $G(\mathbf{r}, \vartheta)$ of light. The intensity autocorrelation, $g_2(\mathbf{r}, \vartheta)$ the 'experimental' measurement, is related to $g_1(\mathbf{r}, \vartheta) \doteq \dfrac{G_1(\mathbf{r}, \vartheta)}{G_1(\mathbf{r}, 0)}$ and has a modulation, $M$, owing to the phase modulation picked up by the specific intensity of light in its passage through the ROI. From $M$ the optical- and mechanical properties of the material in the ROI can be reconstructed. The typical mechanical property is the Young's modulus which influences the amplitude of oscillation of the scattering centres. In an earlier work [8], we have demonstrated the recovery of $p(\mathbf{r})$, the distribution of the mean-squared amplitude of vibration of scattering centres in the object undergoing nearly sinusoidal oscillation under local US forcing (i.e. $p(\mathbf{r}) = \langle |A(\mathbf{r})|^2 \rangle$ where $A(\mathbf{r})$ is the amplitude of vibration, $\langle \rangle$ represents averaging over a volume $(l^*)^3$, with $l^*$ denoting the transport-mean-free path of photons) from $M$. The readily measured quantity in an experiment, $g_2(\mathbf{r}, \vartheta)$ on the boundary of the object, from which $M$ (the experimental measurement) can be computed. (See [26] for the relation between $g_1$ and $g_2$).

Since the local absorption coefficient $(\mu_a(\mathbf{r}))$, $p(\mathbf{r})$ and $n(\mathbf{r})$ influence $M$, it should be possible (at least theoretically) to recover all the above three parameters pertaining to the ROI from $M$. For this, one can make use of the propagation model obeyed by $G(\mathbf{r}, \vartheta)$ in a turbid medium (without US forcing) which is the correlation diffusion equation [25]:

$$-\nabla.\kappa\nabla G(\mathbf{r},\vartheta)+\left(\mu_a+2\mu_s'k_0^2 D_B\vartheta\right)G(r,\vartheta)=S_0(\mathbf{r}_0) \tag{A1}$$

with the boundary condition, $G(\mathbf{r},\vartheta)+\kappa\dfrac{\partial G(\mathbf{r},\vartheta)}{\partial \hat{\mathbf{n}}}=0, \mathbf{r}\in\partial\mathbf{D}$

Here $\kappa$ is the optical diffusion coefficient given as $\kappa=\dfrac{1}{3(\mu_a+\mu_s')}$ where $\mu_a$ and $\mu_s'$ are the optical absorption and (reduced) scattering coefficients respectively. Moreover $D_B$ is the particle diffusion coefficient of the medium, $k_0$ is the modulus of the light propagation vector and $S_0$ is the strength of the isotropic point source at $\mathbf{r}_0$. The term $B(\mathbf{r},\vartheta)=2\mu_s'k_0^2 D_B(\mathbf{r})\vartheta$ is owing to the background temperature induced Brownian motion of the scattering centers. With the focused US beam producing refractive index modulation $(\delta n)$ and oscillations in the scattering centers in the ROI, $G(\mathbf{r},\vartheta)$ is perturbed to $G(\mathbf{r},\vartheta)+G^\delta(\mathbf{r},\vartheta)$. Eqn. (A1) thus becomes

$$-\nabla.\kappa\nabla\left(G+G^\delta\right)(\mathbf{r},\vartheta)+\left(\mu_a+B(\mathbf{r},\vartheta)+A(\vartheta)I_{IR}p(\mathbf{r},\vartheta)\right)\left(G+G^\delta\right)(\mathbf{r},\vartheta)=S_0(\mathbf{r}_0) \tag{A2}$$

The oscillations-induced perturbation is denoted by $A(\vartheta)I_{IR}p(\mathbf{r},\vartheta)$ where $A(\vartheta)=c\sin^2\dfrac{\varpi_a\vartheta}{2}$ and $I_{IR}$ is the characteristic function of the insonified ROI. Here $\varpi_a$ is the acoustic frequency in radians and $c$ is a constant, which depends on $l^*$, $k_a$ (the magnitude of acoustic wave vector) and the elasto-optic coefficient of the material of the object. The boundary condition associated with Eqn. (A2) may be written as

$$\left(G+G^\delta\right)(\mathbf{r},\vartheta)+\kappa\dfrac{\partial\left(G+G^\delta\right)(\mathbf{r},\vartheta)}{\partial\hat{\mathbf{n}}}=0,\ \mathbf{r}\in\partial\mathbf{D} \tag{A3}$$

with $\hat{\mathbf{n}}$ denoting a unit normal to $\partial\mathbf{D}$ at $\mathbf{r}$. The forward problem of UMOT is to solve for $\left(G+G^\delta\right)(\mathbf{r},\vartheta)$ given all the material properties and the US-induced harmonic forcing. This enables evaluating $\mathcal{M}$, the computed approximation to the measurement $M$, through

$$\mathcal{M}(p,\mathbf{r},\varpi_a)\big|_{\mathbf{r}\in\partial\mathbf{D}}=\int_0^\infty G^\delta(\mathbf{r},\vartheta)e^{-j\varpi_a\vartheta}d\vartheta \tag{A4}$$

The oft-used measured quantity $M_1$ is however the modulation depth in $\left(G+G^\delta\right)(\mathbf{r},\vartheta)$ that may be computed via:

$$\mathcal{M}_1(p,\mathbf{r},\varpi_a) = \int_0^\infty (G+G^\delta)(\mathbf{r},\vartheta)e^{-j\varpi_a\vartheta}d\vartheta \tag{A5}$$

Given the measurement, a part of the inverse problem of UMOT is that of the recovery of $p(\mathbf{r})$ given the data $M$ (related to the computable quantity $\mathcal{M}(p,\mathbf{r},\varpi_a)|_{\mathbf{r}\in\partial\mathbf{D}}$), and the forward model introduced through Eqns. (A2) - (A4).

In order to facilitate this inversion, we first rewrite the forward equation as a perturbation equation given by

$$-\nabla.\kappa\nabla G^\delta(\mathbf{r},\vartheta) + (\mu_a + B(\mathbf{r},\vartheta) + A(\vartheta)I_{IR}p(\mathbf{r},\vartheta))G^\delta(\mathbf{r},\vartheta) = -A(\vartheta)I_{IR}p(\mathbf{r},\vartheta)G(\mathbf{r},\vartheta)$$

(A6)

with the boundary condition as

$$G^\delta(\mathbf{r},\vartheta) + \kappa\frac{\partial G^\delta(\mathbf{r},\vartheta)}{\partial\hat{\mathbf{n}}} = 0, \mathbf{r}\in\partial\mathbf{D} \tag{A7}$$

Eqn. (A6) relates $p(\mathbf{r},\vartheta)$ nonlinearly to $G^\delta(\mathbf{r},\vartheta)$ because of the presence of the nonlinear term containing $p(\mathbf{r})$ on the left-hand side (LHS) of it. If we neglect the term containing the product of $G^\delta$ and $p$ from the LHS of Eqn. (A6), then it is computationally more expedient to solve, being linearized in the unknown $p$.

A means of posing the UMOT inverse problem is through the minimization of the following error functional with respect to $p(\mathbf{r})$:

$$\min_{p\in L^\infty(\mathbf{D})} \chi(p) = \frac{1}{2}\|F(p) - M\|^2 + \frac{\beta}{2}\|p\|^2_{L^2(IR)} \tag{A8}$$

Here $M$ is the set of experimental measurements (data) and $F(p) := \mathcal{M}(p,\mathbf{r},\varpi_a)|_{\mathbf{r}\in\partial\mathbf{D}}$ is the operator which takes $p$ as the input and maps it through Eqns. (A6) and (A4) to $M$ (modulo the noise term). The second term of Eqn. (A8) is the regularization term with $\beta > 0$ being the regularization parameter and $IR$ denoting the insonified region. The minimization is usually attempted through a Gauss-Newton (GN) algorithm giving an iterative procedure as $p_{k+1} = p_k - H(p_k)^{-1}G(p_k)$. Here $H$ and $G$ respectively denote the Hessian and gradient of $\chi$. They are approximated by $H(p)(\delta p) = DF^*(p)DF(p)(\delta p) + \beta(\delta p)$ and $G(p) = DF^*(p)(F(p) - M)$. Also, $DF$ denotes the Frèchet derivative of $F$ and $DF^*$ its

adjoint. The details of the implementation of the GN algorithm for UMOT, including the calculation of the gradient and the Hessian via the adjoint of the forward perturbation equation, are given in [8]. We follow the same procedure here to obtain a recovery of $p(\mathbf{r})$ that would be useful as a benchmark for the performance assessment of the proposed method described in Section 2.

The FEM discretization of the PDE in Eqn. (A6) (after linearization) leads to the set of linear algebraic equations represented by

$$\mathbf{K}(\mathbf{p})G^\delta = \mathbf{q} \tag{A9}$$

where $\mathbf{K}(\mathbf{p})$ is the system matrix, with $\mathbf{p}$ denoting the discretized vector for the field $p$, and $\mathbf{q}$ the discretized source vector. Here $G^\delta$ is itself used (with an abuse of notation) to denote the vector of discretized field correlation perturbation. The above discretized form of the perturbation equation has been used in our inversion scheme.